\documentclass[sigconf]{acmart}
\AtBeginDocument{%
  }

\usepackage{verbatim}
\usepackage{multirow}

\usepackage{amssymb}

\usepackage{mathabx}

\usepackage{pifont}
\usepackage{listings}

\usepackage{subcaption}  % For subfigures and captions

\usepackage{algorithm}
\usepackage{algpseudocode}

\newcommand{\FuncCall}[1]{\textbf{\textit{#1}}}
\newcommand{\Continue}{\textbf{continue}}

\newcommand{\True}{\textit{true}}
\newcommand{\False}{\textit{false}}
\newcommand{\COMMENT}[1]{\textit{// #1}}

\usepackage[skip=10pt]{caption}

\lstdefinelanguage{gremlin}{
    morekeywords={g, V, has, out, in, order, by, limit, gt, hasLabel},
    sensitive=true,
    morecomment=[l]{//},
    morestring=[b]"
}

\begin{document}

\title{One-Hop Sub-Query Result Caches for Graph Database Systems}

\author{Hieu Nguyen}
\affiliation{%
  \institution{eBay Inc.}
  \streetaddress{2025 Hamilton Avenue}
  \city{San Jose}
  \state{CA}
  \country{USA}
  \postcode{95125}
}
\email{hieunguyen@ebay.com}

\author{Jun Li}
\affiliation{%
  \institution{eBay Inc.}
  \streetaddress{2025 Hamilton Avenue}
  \city{San Jose}
  \state{CA}
  \country{USA}
  \postcode{95125}
}
\email{junli5@ebay.com}

\author{Shahram Ghandeharizadeh}
\orcid{0000-0001-5109-3700}
\affiliation{%
  \institution{University of Southern California}
  \city{Los Angeles, CA}
  \country{USA}
}
\email{shahram@usc.edu}


\begin{abstract}
This paper introduces a novel one-hop sub-query result cache for processing graph read transactions, gR-Txs, in a graph database system.
%with serial transaction guarantees.
%The cache is implemented inside the graph query processor and transparent to a user.
The one-hop navigation is from a vertex using either its in-coming or out-going edges with selection predicates that filter edges and vertices.
Its cache entry identifies a unique one-hop sub-query (key) and its result set consisting of immutable vertex ids (value).  
When processing a gR-Tx, the query processor identifies its sequence of individual one-hop sub-queries and looks up their results in the cache.
A cache hit fetches less data from the storage manager and eliminates the requirement to process the one-hop sub-query.
A cache miss populates the cache asynchronously and in a transactional manner, maintaining the separation of read and write paths of our transactional storage manager.
A graph read and write transaction, gRW-Tx, identifies the impacted cache entries and either deletes or updates them.
Our implementation of the cache is inside the graph query processing engine and transparent to a user application.
We evaluate the cache using our eCommerce production workload and with rules that re-write graph queries to maximize the performance enhancements observed with the cache.
%further enhance system performance.
Obtained results show the cache enhances 95$^{th}$ and 99$^{th}$ percentile of query response times by at least 2x and 1.63x, respectively.
%with heavy system workloads dominated by either reads or writes.
When combined with query re-writing, the enhancements are at least 2.33x and 4.48x, respectively. 
An interesting result is the significant performance enhancement observed by the indirect beneficiaries of the cache, gRW-Txs and gR-Txs that do not reference one-hop sub-queries.
The cache frees system resources to expedite their processing significantly.
%An interesting result is that the cache enhances the performance of writes and queries that do not reference one-hop sub-queries significantly.
%A cache hit for a sub-query result eliminates the requirement to process the sub-query, freeing CPU, disk, and network resources.
%The system uses these resources to enhance the performance of operations that are not direct beneficiaries of the cache.
%The overhead of our caching solutions is on the write-path of the system when cache misses populate the cache asynchronously in response to misses.

%no processing when compared with executing the sub-query each time.
%This expedites the 95$^{th}$ and 99$^{th}$ percentile of query response times by more than 2x with our production workload.
%We observe a performance improvement for all queries including those that do not use sub-query caching due to reduced load on the graph query processing engine (JanusGraph). 

\end{abstract}


%\ccsdesc[500]{Do Not Use This Code~Generate the Correct Terms for Your Paper}
%\ccsdesc[300]{Do Not Use This Code~Generate the Correct Terms for Your Paper}
%\ccsdesc{Do Not Use This Code~Generate the Correct Terms for Your Paper}
%\ccsdesc[100]{Do Not Use This Code~Generate the Correct Terms for Your Paper}


\maketitle

\section{Introduction}\label{sec:intro}
The graph data model is used by diverse applications, ranging from cybersecurity monitoring to fraud detection, link prediction, and product recommendation~\cite{graph2021,ebay2023,bytegraph2022}.
At eBay, several mission-critical applications rely on our transactional graph database service for their daily critical business operations.
One of our biggest use cases has been in production for several years.  
At the time of this writing, it consists of tens of billions of vertices and edges with a peak traffic in excess of 4000 graph traversals per second.
The workload consists of multi-hop graph read transactions, gR-Txs, and graph read and write transactions, gRW-Txs.
%The gRW-Txs are issued in a batch manner: streaming and scheduled batched loaders.
%The workload consists of multi-hop graph read transactions, gR-Txs, and two types of graph read and write transactions, gRW-Txs, issued in a batch manner: streaming and scheduled batched loaders. 
While the gR-Txs are interactive and their latency impacts real-time decision making, the gRW-Tx are batch updates that do not impact our end-users and hence, favor throughput over latency.

\begin{definition}
A graph read transaction, {\em gR-Tx}, is a Gremlin query that may traverse edges of a graph and apply operations that read the properties of edges and vertices of a graph.
It is a serializable transaction. 
\end{definition}

\begin{definition}
A graph read and write transaction, {\em gRW-Tx}, is a Gremlin expression that may traverse edges of a graph and apply operations that either read or write the properties of edges and vertices of a graph, insert and delete edges and vertices.
It is a serializable transaction. 
\end{definition}

The $50^{th}$ percentile latency of our gR-Txs is competitive, faster than $10$ milliseconds.
However, the $95^{th}$ and $99^{th}$ percentile latency are in excess of 100 milliseconds and 1 second, respectively.
Our objective is to enhance these latencies while preserving ACID semantics of transactions with strict serial schedules.
Our applications do not tolerate either stale or inconsistent query results.

% \begin{itemize}
% \item root : source : start : reference
% \item leaf : destination : adjacent : reachable : neighbor : end : target : qualifying : result
% \end{itemize}

In this paper, we conceptualize a multi-hop gR-Tx as a chain of steps that are evaluated from left to right.
A step may use a boolean predicate to qualify edges that it traverses from a root vertex to produce leaf vertices.
%Some steps examine the property values of an edge from a root vertex and traverse it as long as it qualifies.
%traverse the edges of a root vertex by examining the property values of the edge.
%and examine its properties to produce leaf vertices. 
A subsequent step may apply a boolean predicate to one or more of the properties of these leaf vertices to identify the relevant ones.
% leaf vertices.
%With the resulting leaf vertices, such a step may apply a filter to one or more of their properties to identify the relevant ones.
Such a sequence that traverses an edge and applies predicates is a {\em one-hop sub-query}.
It is characterized by its input being a root vertex id and its output being a set of leaf vertex ids.
%We refer to such a step as a {\em one-hop sub-query} and cache the id of its qualifying leaf vertices.
We cache the result of one-hop sub-queries.
When processing a gR-Tx, we identify its one-hop sub-queries and look up their results in the cache.
If a value is found (a cache {\em hit}), we deserialize the value into a list of leaf vertices and use them to process either the next one-hop sub-query or the remaining portion of the gR-Tx.
Otherwise (a cache {\em miss}), the graph execution engine proceeds to process the one-hop sub-query.  In addition, it initiates an asynchronous transaction that computes the result of the one-hop sub-query and inserts it in the cache for future look ups.

\begin{table*}[!ht]
    \centering
    \begin{tabular}{|c|c|cc||c||cc||c||}
    \hline
    \hline 
         \multirow{2}{*}{Day}&  \multirow{2}{*}{Workload} &  \multicolumn{3}{c||}{95$^{th}$ Percentile Latency (milliseconds)}& \multicolumn{3}{c||}{99$^{th}$ Percentile Latency (milliseconds)}\\ 
         \cline{3-8}
               &  &  $C^-Q^-$ & $C^+Q^+$ & Factor of Improvement & $C^-Q^-$ & $C^+Q^+$ & Factor of Improvement \\
        \hline
        \hline
\multirow{3}{*}{Day 1} & $\widehat R$ & 132 & 43 & 3.07x & 975 & 202 & 4.83x \\
 & $\widehat W$ & 103 & 27 & 3.81x & 732 & 143 & 5.12x \\
 & $\widecheck R$ & 70 & 30 & 2.33x & 676 & 139 & 4.86x \\ 
\hline

%Day 2 & $\widehat R$ & 146	49	2.98	1056	221	4.78 \\
%$\widehat W$ & 109	33	3.30	780	180	4.33 \\ 
%$\widecheck R$ & 78	21	3.71	706	134	5.27 \\ 
\hline

\multirow{3}{*}{Day 4} & $\widehat R$ & 134 & 45 & 2.98x & {\bf 1,020} & 211 & 4.83x \\
 & $\widehat W$ & 100 & 30 & 3.33x & 764 & 166 & 4.60x \\
 & $\widecheck R$ & 82 & 21 & 3.90x & 773 & 138 & 5.60x\\ 
\hline

\multirow{3}{*}{Day 7} & $\widehat R$ & 152 & 48 & 3.17x & {\bf 1,109} & 225 & 4.93x\\
 & $\widehat W$ & 106 & 30 & 3.53x & 775 & 173 & 4.48x\\
 & $\widecheck R$ & 83 & 22 & 3.77x & 762 & 140 & 5.44x\\

\hline \hline

\multirow{3}{*}{Avg} & $\widehat R$ & 137 & 44 & 3.08x & {\bf 1,038} & 212 & 4.90x\\ 
 & $\widehat W$ & 102 & 31 & 3.25x & 770 & 168 & 4.59x\\
 & $\widecheck R$ & 77 & 22 & 3.43x & 727 & 135 & 5.37x\\ 
\hline \hline

    \end{tabular}
    \caption{95$^{th}$ and 99$^{th}$ percentile latency of production workloads on three different days.  Average is across 7 days.}
    \label{tbl:highlight}
\end{table*}

\begin{figure}[!ht]
    \centering
    \includegraphics[width=1\linewidth]{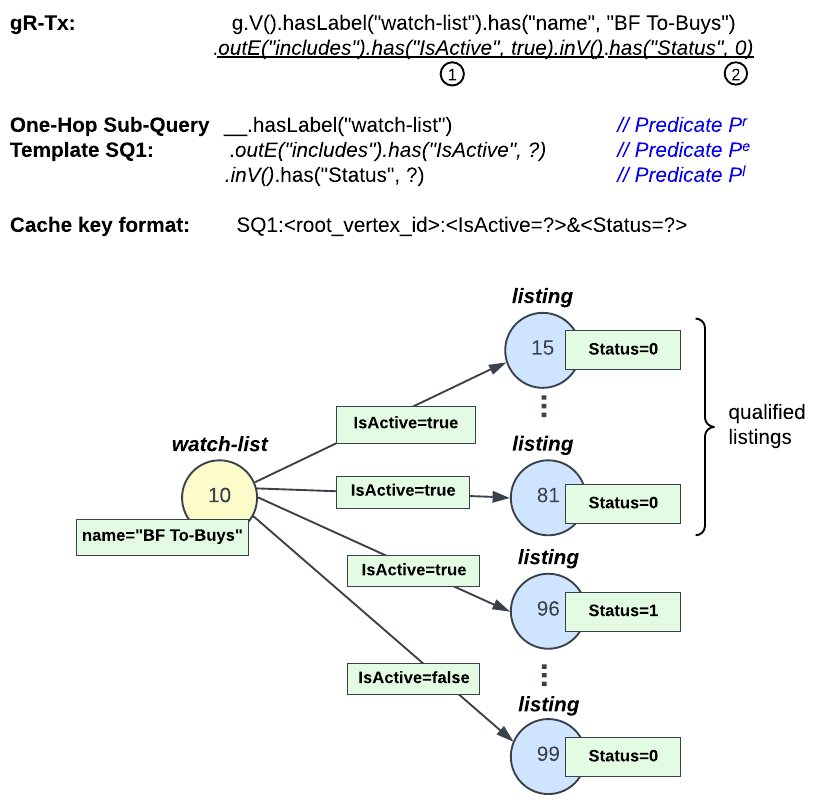}
    \caption{A Gremlin query with a one-hop sub-query.  
    %Shown physical data representations are one possibility out of several.
    }
    \label{fig:sub_query_template_example}
\end{figure}

To illustrate, consider the example Gremlin query of Figure~\ref{fig:sub_query_template_example}.
Its one-hop sub-query is underlined.
Using Vertex 10 in yellow as the root of this sub-query, the sub-query evaluates the ``IsActive" property of the edges outgoing from this vertex to identify those with value equal to true.
The resulting leaf vertices qualify as long as their ``Status" property value equals zero.
We construct a key-value pair where the key identifies the one-hop sub-query instance using its root vertex id and its value is the id of the qualifying leaf vertices.
This is a cache entry and is stored in the cache.
%, see Figure~\ref{fig:sub_query_template_example}.
Every time the query of Figure~\ref{fig:sub_query_template_example} is issued, the query processor constructs the key of its one-hop sub-query and looks up its results in the cache.
If it observes a hit, the query processor uses the obtained value without processing the one-hop sub-query.
In addition to being faster than query processing, a cache hit frees system resources for processing other gR-Txs and gRW-Txs.

Cache misses populate the cache using an asynchronous transaction.
This asynchronous approach enhances the performance of gR-Txs in two ways
when compared with a synchronous approach.
First, it removes populating the cache from the processing path of a gR-Tx.
Second, it prevents a query from becoming a read-write transaction,
utilizing the read path of our storage manager which is faster than its write path.

gRW-Txs may change the state of the graph database, impacting the cached result of one or more one-hop sub-query instances.
%by adding or deleting an edge/vertex or modify the property value of an edge/vertex.(Section~\ref{sec:writes} provides a complete list of possible changes by a write.)
We require gRW-Txs to 
%identify the impacted cache entries and to 
implement either the write-around or the write-through policy~\cite{mrcury2012,dellfluidcache,holland2013flash,sce14,leases2014,writeback2019} to maintain the cache consistent with the graph database.
With write-around, the gRW-Tx deletes the impacted cache entries.
With write-through, the gRW-Tx updates the impacted cache entries.
%A write identifies the impacted one-hop sub-query instance results. These are potential cache entries. The write either invalidates or refills these cache entries.
%Invalidate is a and delete them as a part of the transaction that modifies the graph database.

%Significance of gR-Txs and their 95$^{th}$ and 99$^{th}$ percentile response times. We are eBay. We have demanding user-facing applications with strict response time requirements. We separate our read and write path, processing writes in a batched manner.

%We enhance the response time of reads by caching one-hop sub-query results.  The key insight is that cache lookup for results is faster and more efficient than processing the sub-query.

%How does the cache work? We break a gR-Tx into a sequence of sub-queries. Each sub-query has an input vertex id, identifying a unique vertex of the knowledge graph. The sub-query navigates either the incoming or outgoing edges of the vertex to output a set of vertices. This may examine the value of an edge property.

We evaluated the cache using our production workload across several days.
Our workload includes more than 20 unique gR-Txs with one or more one-hop sub-query traversals.
Only 6 one-hop sub-query templates are required to support all queries.
Some gR-Txs reference one one-hop sub-query while others reference as many as four one-hop sub-queries.
Some gR-Txs reference the same one-hop sub-query template with different parameter values multiple times.

Our evaluation includes use of query re-writing to enhance the efficiency of gR-Txs.
The cache enhances the performance of both the original and re-written gR-Txs.
In this paper, we report on 7 days of gathered results.  Table~\ref{tbl:highlight} presents the 95$^{th}$ and 99$^{th}$ percentile latency for three of those days.
The reported average is across all 7 days.
Table~\ref{tbl:highlight} shows the observed response time of the system with the cache ($C^+$) and query re-writing ($Q^+)$, $C^+Q^+$, when compared with the same system that has disabled these features, $C^-Q^-$.
It also shows the factor of improvement provided by $C^+Q^+$ when compared with $C^-Q^-$.
Our workload is diurnal with heavy$\widehat \_$ and light$\widecheck \_$ system loads dominated by either gR-Txs or gRW-Txs: $\widehat R$, $\widecheck R$, $\widehat W$.
The amount of resources is fixed and identical with and without cache for all workloads.
On the average, the query re-writing combined with the cache provides at least 2.33x and 4.48x improvement in the 95$^{th}$ and 99$^{th}$ percentile latency, respectively.
More importantly, the 1 second 99$^{th}$ percentile latency (highlighted in bold) is now reduced to be lower than a quarter of a second.

The cache combined with query re-writing minimizes the amount of data and processing required by a one-hop sub-query.  
%At the minimum, it reduces two network round trips to process a one-hop sub-query into one to lookup its result in the cache. 
With a %supernode 
vertex that has a high branching factor, both the number of concurrent network invocations and the amount of data transferred are reduced significantly.  
%Our evaluation includes use of query re-writing to enhance the efficiency of gR-Txs.
%The cache enhances the performance of both the original and re-written queries.
%An interesting result is that the cache combined with query re-writing provides a greater benefit than each by itself.  
%Another interesting result are 
%An interesting result is the significant performance gains observed by the indirect beneficiaries of the cache, reads that do not reference a one-hop sub-query and writes that must maintain the cache consistent.
An interesting result is the significant performance gains observed by queries that do not reference a one-hop sub-query and gRW-Txs that must delete cached entries to maintain the cache consistent with the graph database.
The cache frees system resources to expedite their processing significantly.
%enhance their performance significantly.

The {\bf contributions} of this paper include:
\begin{itemize}
    \item Design of a user transparent cache for one-hop sub-queries. gRW-Txs maintain the cache consistent by implementing either the write-around or the write-through policy.
    %either invalidating or refilling the impacted cache entries.  
    Section~\ref{sec:cache}.
    %\item For those queries that reference supernodes, we see a significant improvement in query processing times.  
    \item An implementation that provides strong consistency guarantees. It populates the cache asynchronously and requires gRW-Txs to implement the write-around policy.
    %invalidate cache entries.  
    Both are performed in a transactional manner.  Section~\ref{sec:impl}.

    \item A workflow that enables an administrator to add/remove and activate/deactivate a one-hop sub-query template in an online manner.  Section~\ref{subsec:lifecycle}.

    %\item A workflow that enables an administrator to add and remove a one-hop sub-query template, and activate and deactivate a one-hop sub-query template in an online manner.  Section~\ref{subsec:lifecycle}.

    %\item Query re-writing rules to enhance the performance of gR-Txs.  Section~\ref{sec:rewrite}.
    %\item Incorporated the cache framework into a GraphDB’s query engine. We developed the whole caching framework by modifying the JanusGraph query engine with a caching component with a new caching query step. Query caching can be turned on and turned off. 

    \item Demonstrated the caching framework’s significant performance gain on a large-scale deployment of a graph database in the eCommerce marketplace.
    %, even when the cache hit rate was shown to be not easy to achieve above 90\% hit rate.
    Section~\ref{sec:eval}.
    %\item 
\end{itemize}
A drawback of our cache is its increased writes attributed to cache misses populating the cache.
This results in a higher disk utilization of our storage manager's log servers.  Section~\ref{sec:lesson3}.
%~\ref{sec:eval_coarse}.

The next section presents background, terminology, and a motivating example. 
Subsequently, we present
the above contributions in turn.
%We discuss several interesting features and implementation details of the cache in Section~\ref{sec:discuss}.
Section~\ref{sec:related_work} presents related work. 
Brief conclusions and future research directions are presented in Section~\ref{sec:conclude}.

\section{Background, Terminology, and a Motivating Example}\label{sec:terms}

We assume a directed property graph database consisting of vertices (also known as nodes), directed edges, labels, and properties.
%Typically, vertices represent entities.
%Edges represent connections between vertices and are directed, either going out or into a vertex.
Properties are information added to the vertices and edges.
They are in the form of (attribute, value) pairs.
%Vertices and edges may have labels.
%Labels identify and group similar types of vertices or edges.
Labels categorize and group vertices or edges of similar types.

%A traversal query is a sequence of operations that describe how to traverse the graph from one vertex or edge to another.
%edges to identify relevant to retrieve relevant vertices and edges. 
Figure~\ref{fig:sub_query_template_example} presents an 
%compelling 
example that inspired the development of our caching solution. A watch-list of a user may have multiple listings.
%where each listing may be either active or inactive. 
We model watch-lists and listings as vertices. If a listing is included in a watch-list, there is an edge labeled ``includes'' from the watch-list to the listing. Each listing has a property named ``Status'' with a binary value, either 0 or 1, denoting whether it is available or expired. Moreover, to track a user's behavior, if a listing is excluded from a watch-list, we maintain the edge from the watch-list to the listing and set its ``IsActive'' property value to \texttt{false}. 

A Gremlin traversal query consists of a sequence of operations that describe how to traverse edges of a graph.
An operation may apply a predicate to labels and properties of the vertices and edges.
We assume an operation starts by identifying a root vertex.
It may use predicates to identify its relevant incoming or outgoing edges, pruning the resulting leaf vertices by applying one or more predicates to their labels and properties. 
%They may reference labels and properties of the edges and prune down the resulting vertices by applying a predicate to their labels and properties.
The result is a list of qualifying leaf vertex ids. 
%A traversal query may identify a root vertex by applying a predicate to one or more properties and labels of the vertices.  From this vertex, it may identify its relevant edges, either incoming or outgoing, by examining the property value and label of its edges. This produces a set of leaf vertex ids.
The next operation in the query may use each of these vertex ids as a root vertex and repeat the traversal.
%For each qualifying leaf vertex id, the query may repeat the process of 
%applying a predicate to its property values and those of its edges, producing a new set of leaf vertex ids.   
%The query may navigates the edges of each qualifying vertex id by examining their property value to identify other vertex ids.
This may repeat multiple times until all operations that constitute the query are processed.
The query may use set operators such as union to combine the set of vertices.  
And, with the final set of vertices, it may enumerate a specific property value or apply aggregate functions such as count and group-by~\cite{marko2015}.
The query's final output may be a list of vertices, edges, properties, or any other data extracted from the graph based on the defined operations.

In our example, 
%In Figure~\ref{fig:sub_query_template_example}, 
a watch-list ``BF To-Buys'' consists of 50 listings where 30 edges are active.
A user may be interested in only active listings that have not expired, i.e., Status=0. Assume 25 listings are qualified.
%The user might only interested in active listings and wants to retrieve all active listings in his on their watch-list. 
Figure~\ref{fig:sub_query_template_example} shows the corresponding Gremlin query, a gR-Tx.

We identify the one-hop sub-query instances that constitute a query and cache their resulting vertex ids.
25 vertex ids in our example.
When processing a gR-Tx, we identify its one-hop sub-queries and look up their results in the cache first.  We process a one-hop sub-query only if its result is not found in the cache.  Otherwise, we proceed to process the obtained vertex ids from the cache to continue the next step of the gR-Tx.

\begin{definition}
A {\em one-hop sub-query instance} consumes one root vertex id to output a set of leaf vertex ids.
It does not change the state of the graph database.
It may navigate either the incoming or outgoing edges of the root vertex by applying a predicate to either the property values of its edges, their labels, or both.
It may qualify a leaf vertex by applying one or more boolean predicates to its property values, labels, or both.
The leaf vertex ids are the output of a one-hop sub-query.
Their number is the cardinality of the sub-query.
It may range from zero (output is the empty set) to a large value (output of a supernode with many qualifying edges and leaf vertices).

%A {\em sub-query} inputs one vertex id to output a set of vertex ids. It computes its output by examining the value of properties associated with its input vertex and its edges. The number of vertex ids in the output is the cardinality of the sub-query. It may be zero (output is the empty set) to a large value (output of a super vertex with many qualifying edges). 
\end{definition}

\begin{definition}
A {\em one-hop sub-query template} identifies the structure of a sub-query instance that is cached.
%It consists of a set of predicates that qualify a root vertex {P$^r$}, an edge {P$^e$}, and a leaf vertex {P$^l$}.
It consists of a predicate P$^r$ to qualify a root vertex, a predicate P$^e$ to qualify an edge, and a predicate P$^l$ to qualify a leaf vertex.
A predicate may be empty, returning true always.
%In addition to values, 
It may use a wildcard `?'.
% on its predicates.
%as the value used by an instance.
A wildcard qualifies all possible values as a match.
% for a predicate.
It is possible for a sub-query template to use a value instead of a wildcard.
In this case, only sub-query instances with the matching value are cached.
\end{definition}
A one-hop sub-query instance must match a one-hop sub-query template in order for its results to be cached.
Otherwise, it is not cached and gRW-Txs will not delete or update its cache entry.
The use of a value in a one-hop sub-query template is justified when a large percentage of gR-Txs use the value.

With the example query of Figure 1, a graph query execution engine such as JanusGraph processes the query by issuing 2 sequential and $n$ concurrent requests to a storage manager.  A cache hit reduces these to 2 requests.  To elaborate, the two sequential requests by the JanusGraph are (1) retrieve the id of the watch-list vertex using an index, and (2) a range selection predicate to retrieve all edges and their property values, see \textcircled{1} in Figure~\ref{fig:sub_query_template_example}.  JanusGraph performs local processing to filter those edges that satisfy the predicate "IsActive=true", identifying $n$ qualifying edges.  In turn, these edges identify listing encoded as a part of the edge ids.  JanusGraph issues $n$ concurrent range selection predicates to retrieve the ``Status'' property of each listing, filtering those that satisfy the predicate ``Status=0'', see \textcircled{2} in Figure~\ref{fig:sub_query_template_example}.
We assumed 25 listings satisfy this predicate and their vertex ids are returned to the client.

The template for the one-hop sub-query is:
%Caching sub-query instances is a good fit for this example.
%Caching instances of sub-queries works well in this case.
%A sub-query template for caching fits well in this case. 
%The sub-query template
%Its template is:
\begin{lstlisting}
__.hasLabel("watch-list")              // Root predicate
  .outE("includes").has("IsActive",?)  // Edge predicate
  .inV().has("Status",?)               // Leaf predicate
\end{lstlisting}
A key for its instance (see underlined portion of gR-Tx in Figure~\ref{fig:sub_query_template_example}) might be 
%of this template is 
SQ1:10:IsActive=true\&Status=0 where SQ1 is the template identifier, 10 is the immutable id of the JanusGraph vertex of the watch-list with ``name" equal to ``BF To-Buys", and IsActive=true\&Status=0 are the parameters provided by the edge and leaf predicates of the instance, see gR-Tx in Figure~\ref{fig:sub_query_template_example}. The predicate of gR-Tx identifies the relevant root vertices and is not required in the cache key.
Its value consists of 25 vertex ids serialized into an array of bytes.

%Every time the one-hop sub-query observes a cache hit, the 31 requests\footnote{The index lookup for the qualifying vertex is required to lookup the cache key.} are reduced to one for a cache lookup.
A cache hit reduces $n+2$ requests to 2.
The JanusGraph is still required to retrieve the immutable id of the watch-list vertex using an index.
It uses this id to construct the key for the cache entry and looks up its value in the cache.
%one lookup for the value of a key.
A hit obtains the final result, i.e., 25 vertex ids, reducing the amount of data retrieved and eliminating the local processing by JanusGraph.

%and the cost of de-serializing a value and local processing (for evaluating the query predicate) is eliminated.
%also reduced significantly.
%With a cache miss, the processing with two network round-trips of getRange requests incurs the additional network round-trip of cache lookup.

A gR-Tx may consist of multiple one-hop sub-query instances.
%A well-chosen sub-query template can also be used in multiple query templates. 
An example is, given a listing id ``L1'',
%For example, another query is given a listing id ``L1'', 
retrieve all other active listings that are included in at least one common watch-list with ``L1''.
%Its expression in Gremlin is
Its Gremlin query is:

\begin{lstlisting}
g.V().hasLabel("listing").has("id", "L1")
  // find all watch-lists that include this listing
  .inE("includes").has("IsActive",true).outV()
    .hasLabel("watch-list")
  // for each watch-list, find all listings include it
  .outE("includes").has("IsActive", true).inV()
    .has("Status", 0)
  .valueMap()
\end{lstlisting}
Its two one-hop sub-query templates\footnote{\color{blue}{has}\color{black}(\color{red}"id", "L1"\color{black}) from the root predicate is not required in Template 1.  This makes the template applicable to a query with a different listing id than ``L1''.} are:
\begin{lstlisting}[numbers=left]
Template 1: __.hasLabel("listing").inE("includes").has("IsActive",?).outV().hasLabel("watch-list")
Template 2: __.hasLabel("watch-list").outE("includes").has("IsActive",?).inV().has("Status", ?)
\end{lstlisting}
An instantiation of the first one-hop sub-query template produces a set of leaf vertex ids.
Each vertex id is used as the root vertex of the second one-hop sub-query instance to reference a cache entry.

%For each such vertex id, 
%The second one-hop traversal of the query matches the defined sub-query template and may reference multiple cache entries, where each cache key is associated with each watch-list vertex found at the first-hop traversal.

%}

\begin{definition}
A {\em supernode} is a vertex with many in-coming, out-going, or a combination of in-coming and out-going edges.  
\end{definition}

A supernode may either be a root or a leaf vertex of a one-hop sub-query instance.
When a supernode is the root of a one-hop sub-query, it may result in many traversals to compute the result of the sub-query.  However, it may not produce many leaf vertex ids.
When a supernode is a leaf vertex, it means the supernode appears in the result of many one-hop sub-query instances.
Thus, a gRW-Tx that impacts a supernode may either delete or update many cache entries.

\section{One-Hop Sub-Query Result Cache}\label{sec:cache}
This section provides a conceptual description of the one-hop sub-query result cache.
These concepts may be implemented in a variety of ways.
We detail an implementation in Section~\ref{sec:impl}.
This implementation is a subset of the concepts presented in this section.
For example, it implements the write-around policy only while this section presents both the write-around and the write-through policies.
%the concept of both invalidate and refill.

A cache entry is a key-value pair that corresponds to the result of a one-hop sub-query instance.
The key identifies a unique one-hop sub-query instance.
Its value is the output of the one-hop sub-query instance, a list of leaf vertex ids.
We construct the key using (a) the identifier of the one-hop sub-query template of the instance, (b) its input root vertex id, 
%(b) the vertex property names and values referenced by the predicates (if any), 
(b) the values identified by the properties with wildcards in $P^{e}$, and (c) the values identified by the properties with wildcards in $P^{l}$. 
%(b) the direction of its referenced edge, edge label, edge property names and values referenced by $P^{e}$, (c) the leaf label, property names and values referenced by $P^{l}$.
The one-hop sub-query template must be registered with the system in order for its instances to be cached.
We maintain 
a list of 
$T$ templates.
%predicates referenced by a sub-query. 
gRW-Txs use this list to identify cached keys that should either be deleted or updated.
%invalidated or refilled.

\subsection{Processing gR-Txs}

When processing a gR-Tx, the system identifies its list of one-hop sub-query instances and the sequence in which they should be processed.
For a sub-query instance and its root vertex id, it constructs the key of its cached entry.
It looks up the key in the cache.
If the cache produces a value, the query processor (e.g., JanusGraph) has observed a cache hit.
It deserializes the value to obtain a list of leaf vertices.
It uses each resulting vertex id as the root vertex of the next sub-query.
%and enumerates the set of leaf vertices that constitute the root vertex to process the next sub-query.
With a cache miss, the query processor executes the one-hop sub-query instance.
%It inserts the sub-query instance and the root vertex id into a queue that is processed by background threads.
It uses the resulting leaf vertex ids to process the next sub-query instance.
It may populate the cache either synchronously or asynchronously.
It repeats this sequential process until it exhausts all the sub-query instances that constitute the gR-Tx.
It processes the final set of vertices using the final clause of the gR-Tx.
At this point, it has completed processing the gR-Tx and produces the final result set as the output.

%We realize this by populating the cache asynchronously using the CP threads.

We assume vertex ids used to construct the cache keys and their values are immutable. 
In addition, we assume the label of a vertex or an edge is immutable.

%Immutable vertex ids used to construct the cache key and the cached values.
%A cached value consists of a list of vertex ids.
%One or more properties of a vertex may be immutable.
%For example, with JanusGraph, the Label property of a vertex is immutable.

\subsection{Processing gRW-Txs}\label{sec:writes}
A gRW-Tx identifies the impacted cache entries and implements either the write-around or the write-through policy.
While the write-around policy deletes the impacted cache entries, the write-through policy updates them.
%impacted cache entries in a transactional manner. 
%Performing these operations in a transactional manner means the invalidation and refill operations 
These are performed as a part of the 
%write transaction,
gRW-Tx, preserving the consistency of the cache entries with the database state.
A gRW-Tx may consist of one or more of these changes:
\begin{enumerate}
    \item add/delete a vertex $V_{i}$,
    \item add/update/delete a property %with a value to 
    of a vertex $V_{i}$, 
    \item add/delete an edge from vertex $V_i$ to vertex $V_j$,
    %between two vertices, $V_{i}$ and $V_{j}$, 
    \item add/update/delete a property %value 
    of an edge between two vertices, $V_{i}$ and $V_{j}$.
    %\item perform several of the above as one transaction.
\end{enumerate}

Adding a vertex impacts no cache entry.
A new vertex added alone with no edges is isolated from the graph. 
It does not impact the result of a one-hop sub-query. 
The remaining changes may impact one or multiple cache entries. 
%With changes on an edge, the direction of the edge is important for identifying the impacted cache entries.

%Refill has two possible definitions when 
A change by a gRW-Tx may produce a new cache entry that did not exist previously.
An example is a gRW-Tx that adds a new property value to $V_i$, causing some one-hop sub-queries to produce a vertex list for $V_i$ that did not exist previously.
In such scenarios, there are two ways to implement write-through.
A ``lazy'' write-through does nothing.
%One definition of refill is to do nothing.
%This will require a future gR-Tx that references the one-hop sub-query with $V_i$ to observe a miss and populate the cache.
%, causing a CP to compute the missing cache entry and populate the cache asynchronously.
A ``pro-active'' write-through requires 
%Another definition is to require 
the gRW-Tx to populate the cache with the new entries, enabling the future gR-Tx to observe a hit.
The former is advantageous when there exists no future gR-Txs that reference the new cache entry.
The latter is advantageous when many future gR-Txs reference the new cache entry almost immediately and frequently. 
Its disadvantage is that it slows down the gRW-Txs by requiring them to populate the cache.
This section assumes the lazy write-through.

%In order for a write to process sub-queries to identify the impacted cache entries, 
%it may require the vertex property names and values used by the different sub-queries.
%The property names may be obtained once at system initialization time by processing the sub-query templates.
%The value of these properties for the impacted vertices may be obtained once before the write identifies the impacted sub-queries.

%A change incurs two additional steps: (1) identifies impacted cache keys and (2) invalidates/refills them.
%A cache enabled graph store may traverse and fetch property names and values of related vertices and edges to identify the impacted cached keys.  While a change on a vertex requires both $P^{r}$ and $P^l$ of a sub-query template to be re-evaluated, a change on an edge requires $P^e$ to be re-evaluated.

A change computes its impacted keys by analyzing the $T$ one-hop sub-query templates $SQ_t$, $1 \leq t \leq T$.  Write-around deletes these keys.  Write-through may either delete an impacted key $k$ or append/remove a vertex id from the value $v$ of a key $k$.  With the latter, a value $v$ must exist in the cache for the key $k$.  Otherwise, no vertex id is added or removed from $v$.
In the following, we assume the impacted $k$-$v$ exists in the cache.

%With invalidation, a change identifies a list of cached keys to delete.
%With refill, except in cases where a root vertex is deleted (which requires its corresponding cached keys to be deleted), cached values are updated. Each change identifies a list of candidates \{ k, S, V \} where k is the cached key, S indicates 
%the state that 
%either old or new, and V is a leaf vertex id. The cached value $v$ is obtained by looking up $k$ in the cache.
%$v$ is a list of leaf vertex ids.
%If S is old, V is deleted from $v$. If S is new, V is added to $v$.

%Below, we detail each change and how the cached keys (with invalidation) or the list of candidates (with refill) are identified.

{\bf Delete a vertex $V_i$:}
For each one-hop sub-query template $SQ_t$, its $P^r$ is evaluated on $V_i$. If $V_i$ qualifies then it might be a root vertex id of a cached instance of the template.
The change computes the impacted cached keys by instantiating 
%$P^r$'s wildcards using $V_i$'s property values, and 
$P^e$'s wildcards using the property of edges from $V_i$, and $P^l$'s wild cards using the property of leaf vertices reachable from the qualifying edges of $V_i$.
%It instantiates the query template's wildcards using $V_i$'s property values to obtain its query instance.
%It constructs a query instance that is based on the one-hop sub-query template to determine the values of the wildcard properties.
%We execute the query instance to obtain values that determine the one-hop sub-query instances that share $V_i$ as the root vertex.
%We use $V_i$ in combination with these values to identify the impacted cached keys, \{ k \}. 
Both write-around and write-through delete these keys.
%To find all possible values of properties with wildcards, it constructs a query instance that is based on the one-hop sub-query template by instantiating the values of the wildcard properties.
%without the wildcard properties.
%These values determine the one-hop sub-query instances that share $V_i$ as the root vertex. 
%From $V_i$ and these values, cache keys are identified and deleted regardless of invalidation or refill.
%In an ordered key-value store, instead of querying, all cached keys for a root vertex can be deleted by a clear range operation, see Section~\ref{sec:impl}.

Next, $P^l$ of each one-hop sub-query template $SQ_t$ is evaluated on $V_i$. If $V_i$ qualifies then it might be a leaf vertex of $SQ_t$.
The change identifies all root vertices $V_j$ that qualify $V_i$ as a leaf vertex with $SQ_t$. It does so by constructing another one-hop sub-query to traverse back where the edge direction is the reverse of the $SQ_t$. This query identifies all possible root vertices by evaluating them and their corresponding edges with $P^r$ and $P^e$. 
With each root vertex $V_j$, it computes the corresponding cached key $k$ with the property values in the sub-query template and the values of properties with wildcards obtained from $V_i$, $V_j$ and the qualifying edges between them.
%Each is a candidate \{ k, S, $V_i$ \}. 
%S is set to old because the evaluations are applied prior to $V_i$ being deleted.
Write-around deletes these keys. 
Write-through removes $V_i$ from the value of each key. 
%S is always old 
%when $V_i$ still exists.

{\it Example 1: } Consider a gRW-Tx that deletes the watch-list Vertex 10 in Figure~\ref{fig:sub_query_template_example}. This vertex qualifies $P^r$ of SQ1 template. 
Querying outgoing edges and the corresponding vertices computes the values for ``IsActive'' and ``Status'' that identify the impacted cached keys:
$\{$SQ1:10:IsActive=true\&Status=0,
SQ1:10:IsActive=true\&Status=1,
SQ1:10:IsActive=false\&Status=0,
SQ1:10:IsActive=false\&Status=1$\}$.
%of the one-hop sub-query template SQ1.
Both write-around and write-through delete these keys.

% Instead of SQ1, assume we have SQ2 with no wildcard
% \begin{lstlisting}
% __.hasLabel("watch-list").outE("includes").has("IsActive",true).inV().has("Status", 0)
% \end{lstlisting}
% only one cache key is deleted. 

{\it Example 2: } Assume a gRW-Tx deletes the listing with vertex id 15 in Figure~\ref{fig:sub_query_template_example}. The vertex has the property ``Status'' referenced by P$^l$ of the SQ1 template.
We identify the edges that are incoming to this vertex and their root vertices.
This identifies the root vertex id 10.
Using the predicate P$^e$ of the SQ1 template, the ``IsActive'' property value of the edge identifies the impacted cached key:
SQ1:10:IsActive=true\&Status=0. 
Write-around deletes this key. 
Write-through removes Vertex 15 from the value $v$ (list of leaf vertices) identified by the key.
In our example of Section~\ref{sec:terms}, write-through reduces the 
%number of vertex ids identified by 
cardinality of $v$ from 25 to 24 vertex ids.

{\bf Add/Update/Delete a property of a vertex $V_i$:}
The property must be referenced by a predicate of one-hop sub-query template to impact a cache entry.  Otherwise, the change is applied to the graph store with no modifications to the cache entries.

For each one-hop sub-query template $SQ_t$ with a predicate P$^r$ or P$^l$ that references the property, denote the old and the new value of the property as $\Omega_{old}$ and $\Omega_{new}$, respectively.
If the change is deleting (adding) the property then $\Omega_{old}$ ($\Omega_{new}$) is considered.
A change that updates the property value considers both $\Omega_{old}$ and $\Omega_{new}$.
The change computes the impacted keys using the discussion of \emph{Delete a vertex V$_i$}.
Write-around deletes these keys.
Write-through removes the vertex $V_i$ from those cache entries computed using 
$\Omega_{old}$ and adds $V_i$ to those keys identified by $\Omega_{new}$.
%If the property is referenced by P$^r$ or P$^l$ then denote V$_i^{old}$ as the old value of V$_i$, and V$_i^{new}$ as its new value. 
%If the property is added or deleted then only V$_i^{new}$ or V$_i^{old}$ is considered, respectively. 
%If it is updated then both V$_i^{old}$ and V$_i^{new}$ are considered. 
%We compute V$_i^{old}$ and V$_i^{new}$ in the same way as V$_i$, as described in \emph{Delete a vertex V$_i$}, with one difference. With refill, the state S in the list of candidates is \emph{old} for V$_i^{old}$ and \emph{new} for V$_i^{new}$.

{\it Example 3: } 
Consider an update that changes the ``Status'' property of the listing vertex with ID 15 from 0 to 1, i.e., $\Omega_{old}$=0, $\Omega_{new}$=1. This property is referenced by P$^l$ of SQ1.
%We use its old and new value to compute the cached keys SQ1:10:IsActive=0\&Status=0 and SQ1:10:IsActive=0\&Status=1.
We use its old and new value to compute the impacted keys
%\{$SQ1:10:IsActive=0\&Status=0, old, 15$\}$ and $\{$SQ1:10:IsActive=0\&Stat us=1, new, 15$\}$.
SQ1:10:IsActive=0\&Status=0, and SQ1:10:IsActive=0\&Status=1.
Write-around deletes both keys. 
Write-through removes 15 from the value of the key identified using $\Omega_{old}$ and adds 15 to the value of the key identified using $\Omega_{new}$.

{\bf Add/Delete an edge between two vertices, $V_i$ and $V_j$:}
The direction of the edge is important. 
We assume the edge is outgoing from $V_i$ to $V_j$.
Its inverse is a simple switch of subscripts in the following paragraphs.

For each one-hop sub-query template $SQ_t$, it may traverse an edge in either \emph{outgoing}, \emph{incoming}, or \emph{both} directions.
%The sub-query template direction may be \emph{outgoing}, \emph{incoming}, or \emph{both}. In the first case, only one possible (root, leaf) pair is ($V_i$, $V_j$). The reserve applies to the second case, which is ($V_j$, $V_i$). With the third, both are possible.
The first two compute one cached key.
The third computes two cached keys.
%One cached key is identified by each of the first two.  The third identifies two cached keys.
Below, we consider a $SQ_t$ that traverses an {\em outgoing} edge.
The others are a trivial extension.
%At most two cached keys are identified for the third case.

The change computes the impacted keys by using $SQ_t$'s $P^e$ on the edge, and its $P^r$ and $P^l$ using $V_i$ and $V_j$, respectively.
Write-around deletes these keys.
If the outgoing edge is added (deleted), write-through adds (removes) $V_j$ to (from) the value of the key.

%identify the list of candidates \{ k, S, V \} (for refill). $V$ is either $V_i$ or $V_j$ depending on which is the leaf vertex. S is old if the edge is deleted and new if the edge is added.  Invalidate deletes these keys.  Refill either adds or deletes the identified vertex id to the value of the candidate key using the value of S.

{\it Example 4: } Consider a gRW-Tx that adds an {\em outgoing} edge "includes" from the watch-list 10 to a listing id 105 with Status=0 in\footnote{Listing 105 is not shown in Figure~\ref{fig:sub_query_template_example}.} Figure~\ref{fig:sub_query_template_example}.
Using the predicates of SQ1, this change computes the key SQ1:10:IsActive=true \&Status=0.
%It computes the candidate $\{$SQ1:10:IsActive=true \&Status=0, new, 105$\}$.
%It impacts the cached key SQ1:10:IsActive=true\&Status=0.
Write-around deletes this key.
Write-through appends 105 to the value of this key. 

{\bf Add/Update/Delete a property of an edge $E$ between two vertices, $V_i$ and $V_j$:}
The property must be referenced by a predicate $P^e$ of a one-hop sub-query template to impact a cache entry.  Assuming this is the case\footnote{If not, the change is applied to the graph store with no modification to the cache entries.}, the discussion of Add/Delete an edge applies.
For each one-hop sub-query template $SQ_t$ with a $P^e$ that references the property, adding a property to an edge is logically equivalent to adding an edge with the new property value while maintaining the other property values associated with the edge.
Similarly, 
removing a property is logically equivalent to removing an edge.
%while maintaining the value of other properties associated with the edge.
Updating the value of an edge property is the same as deleting an edge with the old value and adding a new edge with the new value.% (while maintaining the value of other properties associated with the edge).

%Denote $E^{old}$ the old state of $E$, and $E^{new}$ its new state.  Similar to the discussion of Add/Delete an edge,  When either adding or deleting an edge property, we consider either E$^{new}$ or E$^{old}$, respectively.  When updating the edge property value, we consider both E$^{new}$ and E$^{old}$.
%Also similar to a property change on a vertex, only E$^{new}$ or E$^{old}$ is considered when adding/deleting a property. Both are considered when updating a property.

%Each of E$^{old}$ or E$^{new}$ is examined as described in the discussion of adding/deleting an edge between two vertices. Each state may identify at most two candidates
%impacted keys,
%(for invalidation) or two keys for refill 
%\{ k, S, V \}. $V$ is either $V_i$ or $V_j$ depending on which is the leaf vertex. S is old if the state being examined is $E^{old}$ and new with E$^{new}$. 

{\it Example 5: } Consider a gRW-Tx that updates the property ``IsActive'' of the outgoing edge from the watch-list 10 to the listing id 15 from true to false.
Logically, this is the same as deleting the outgoing edge with ``IsActive=true'' from the watch-list 10 to the listing id 15 and adding an outgoing edge with ``IsActive=false'' from the watch-list 10 to the listing id 15.
The first computes key SQ1:10:IsActive=true
\&Status=0.
Write-around deletes this key while write-through removes 15 from the value of this key.
The second computes the key SQ1:10:IsActive=false
\&Status=0.
Write-around deletes this key while write-through appends 15 to its value.
%the value of this key.

%It computes two candidates:
%$\{$SQ1:10:IsActive=true\&Status=0, old, 15$\}$ and $\{$SQ1:10:IsActive=false\&Status=0, new, 15$\}$.
%Two impacted cache keys are SQ1:10:IsActive=false\&Status=0 and SQ1:10:IsActive=true\&Status=0. Invalidation deletes both keys.
%Invalidate deletes both keys.  Refill removes 15 from the value of the first key and appends 15 to the value of the second key.
%the cache values are updated to include 15 for the first and exclude 15 for the second.

\section{An Implementation}\label{sec:impl}
\begin{figure}[!ht]
    \centering
    \includegraphics[width=0.7\linewidth]{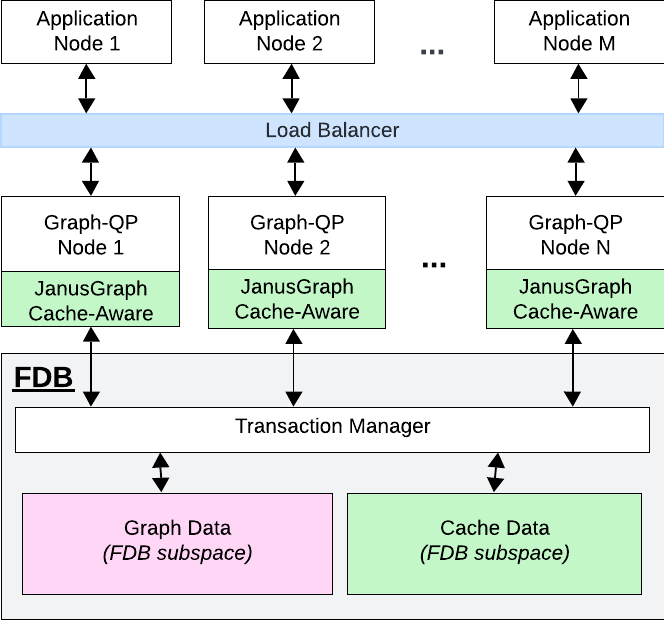}
    \caption{Graph service architecture.}
    \label{fig:arch}
\end{figure}
 
%We have implemented the sub-query caching framework in our system (see Figure~\ref{fig:arch}).
We implemented the one-hop sub-query caching framework with the write-around policy using JanusGraph~\cite{janusgraph} as our graph query processor (Graph-QP), a transactional storage managed named FoundationDB (FDB)~\cite{fdb2022}, and a Service Coordinator (SC).  
%This system consists of Graph Query Processors (Graph-QPs), a transactional storage manager named FoundationDB (FDB)~\cite{fdb2022}, and a Service Coordinator (SC).
FDB consists of stateful components that implement both the graph database and the cache as different subspaces.
The Graph-QPs are stateless virtual servers that are created on demand.
They process gR-Txs by looking up the cache for the result of one-hop sub-queries, and implement the write-around policy with gRW-Txs.
%process gRW-Txs using cache invalidate.
A newly launched Graph-QP registers itself with SC. SC facilitates Graph-QP discovery and removes bad\footnote{Unreachable, failed, or slow.} Graph-QPs from a deployment.
%, and dynamically re-sizes a deployment based on the imposed workload.  
It also implements the life-cycle of a sub-query template from the time a system administrator registers it to the time the administrator disables and removes it from the system, see Section~\ref{subsec:lifecycle} for details. 

A cache entry is a key-value pair stored in FDB.
The value of a key is the result of its corresponding one-hop sub-query instance, i.e., a list of leaf vertex ids.
There are multiple physical organizations of the cache entries using FDB.  This section describes one out of several possible designs.  With this design, the key consists of three parts:
a prefix, a root vertex id, and the parameter values of the predicates presented in its one-hop sub-query instance.
The prefix is unique for each one-hop sub-query template and may be a generic byte array.
However, we use unique strings for the prefix to make keys and monitoring logs easier to process for debugging and troubleshooting purposes.
All cached keys belonging to a sub-query template have the same unique prefix.
FDB as an ordered key-value store groups these keys together, deleting the keys belonging to a sub-query template quickly and efficiently\footnote{Such a delete operation is required when removing a cache template, see Section~\ref{subsec:lifecycle}.}.

%{\color{blue}
Cache values are compressed to reduce cache storage and network transmission overhead. We use Zstd~\cite{Zstd} in our implementation, which in our micro-benchmark shows advantages in both compression/decompression time and compression ratio compared with others. Even with compression, it is possible that a cached value may become larger than the allowed FDB max value size (100k bytes). To handle this case, Graph-QP splits the value into chunks, identifying Chunk $i$ by appending the original key with $i$.
%and stores each chunk as a cache value where its cache key is the original key appended with the chunk index.
%Invalidating 
Deleting a cached entry is a clear range with the cache key as the prefix of the range.
Retrieving a cached entry becomes a range query on its key as the prefix to retrieve all the chunks. Chunks are combined to construct the original compressed value.
Next, this value is decompressed and de-serialized to obtain the cached sub-query result. %Similarly, deleting a cache key becomes clearing a range with the cache key as the prefix of the range.
%}

\begin{figure}[!ht]
    \centering
    \includegraphics[width=0.8\linewidth]{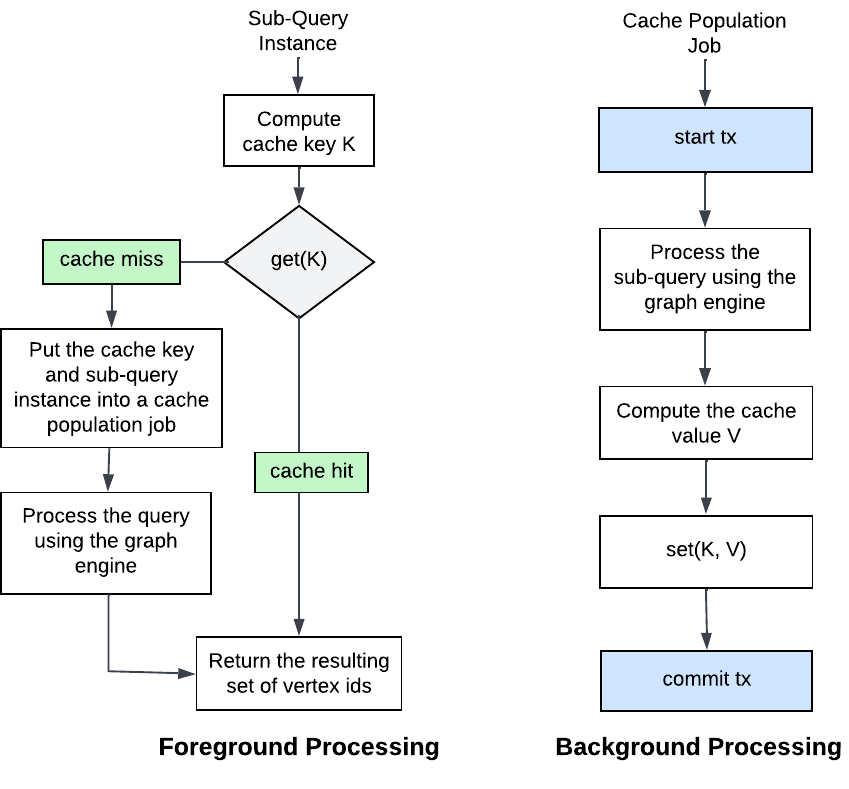}
    \caption{Asynchronous read query processing with cache.}
    \label{fig:async_read_processing}
\end{figure}

Our implementation populates the cache asynchronously as follows.
%once a one-hop sub-query instance observes a cache miss.
%This expedites the processing of a query by preventing it from incurring the overhead of populating the cache.
%Moreover, it prevents a read query from becoming a read-write transaction.
%In our implementation, a sub-query instance that observes a cache miss populates the cache asynchronously.
%With a cache miss, the query processor executes the sub-query instance.
Once the query processor observes a cache miss for a one-hop sub-query instance and a root vertex id, it inserts both in a queue that is processed by a background {\em Cache Populate} (CP) thread.
Next, it processes the sub-query instance using the root vertex id.

The CP threads are hosted on the Graph-QPs of Figure~\ref{fig:arch}.
%One or more background {\em Cache Populate} (CP) threads process the queue.
% of one-hop sub-query instances and their root vertex ids.
%These threads are 
A CP thread executes a transaction that consists of the processing of the one-hop sub-query instance using its provided root vertex id, constructing the unique key $k$ for this one-hop sub-query instance and its vertex id, obtaining the resulting leaf vertex ids and serializing them as the value $v$ of the key, and inserting this $k$-$v$ pair in the cache.
One may view this as a gRW-Tx.
This transaction may fail to commit.
The CP thread executes the entire transaction a pre-specified number of times prior to discarding it. 

Our implementation executes a one-hop sub-query that observes a cache miss twice\footnote{We investigated other designs that execute the sub-query once by increasing the query execution time.  This proved undesirable because it slowed down gR-Txs.
Our goal is to execute gR-Txs as fast as possible.}.
Once by the query processor (JanusGraph) and a second time by a CP thread using the same query processor (JanusGraph).
This prevents a read query from becoming a read-write transaction.
To elaborate, FDB differentiates between its read and write paths, optimizing each independent of the other.
%FDB's read path allows a client to directly retrieve data from a globally consistent snapshot version from a FDB storage manager process.
Its read path allows a client to access data from a globally consistent snapshot version directly from an FDB storage manager process.
%For example, FoundationDB implements an efficient read path by allowing a client to fetch its data from a global consistent snapshot version directly from a storage manager process.
Its write path implements the optimistic concurrency control protocol with multi-version concurrency control.
In essence, a gR-Tx that fetches a large amount of data does not incur the overhead of a concurrency control protocol and transaction commit.  
We ensure gR-Txs that observe a cache miss use the FDB read path while CPs that populate the cache with the missing cache entries use the FDB write path.

We implemented the write-around policy
of Section~\ref{sec:writes} in Graph-QP.
It identifies the cached keys impacted by a gRW-Tx and deletes them.
It uses a Gremlin-provided mutation listener interface to process a change.
%{\color{blue}
%To handle invalidation, we modified the write path to identify impacted cache keys. Gremlin provides a mutation listener interface that raises events when a change occurs. For each change, we capture the old and new state, and check whether either state impacts a sub-query template. 
%To handle gRW-Txs in the presence of the cache, we modified the Graph-QP to identify impacted cache keys and invalidate them. Gremlin provides a mutation listener interface that can be implemented to process a change. 
For each type of change, it captures the old and new states, and for each one-hop sub-query template, it evaluates whether either state impacts cache keys.
%For example, if the sub-query template Q has a predicate Status=0, a change to update status from 1 to 0 has the new state impacts Q, while a change to update status from 2 to 2 does not impact Q. If another sub-query template has Status=?, a change to update status from 1 to 2 has both the old state and new state impact Q.
Appendix~\ref{sec:pseudocode} provides the pseudo-code to implement each change.
To illustrate, consider a sub-query template Q with ``Status=?''.
An update that changes the ``Status'' property value of a vertex from 1 to 2 invokes a listener that provides both the old and new value of ``Status''.
We use this information per discussion of Section~\ref{sec:writes} to construct the key of the impacted cache entries and delete them. 
With FDB, a write transaction buffers its writes at the client.
Thus, both the deleted cached keys and the changes to graph data are buffered and submitted together at transaction commit time.

%With FDB, a write transaction buffers its writes and submits them at its commit time. Thus, both the deleted cached keys and the changes to graph data are buffered and submitted together at transaction commit time.
%invalidation of cached keys and other graph changes are buffered in a write transaction at the FDB client and are submitted at its commit time.

\begin{table}[!ht]

    \centering
    \begin{small}
    \begin{tabular}{|l|c|}
        \hline
    \multirow{2}{*}{Write Change Type} & Max keys deleted/ \\
    & ranges cleared \\ \hline
Add/Delete an edge & Delete 2 keys \\
%Delete an edge & Delete 2 keys \\
Add/Delete an edge property & Delete 2 keys \\
Update an edge property & Delete 4 keys \\
Delete an existing root vertex & Clear 1 range \\
Add/Update/Delete a root vertex property & Clear 1 range \\
Delete an existing leaf vertex & Delete L keys \\
Update a leaf vertex property & Delete 2L keys \\
\hline

    \end{tabular}
    \end{small}
    \caption{Different types of writes and their number of impacted keys and key ranges (per sub-query template).  $L$ is the number of leaf vertices.}
    \label{tbl:writeInval}
\end{table}

% \vspace{-10pt} 

With our implementation, a gRW-Tx either deletes a fixed number of keys or clears a fixed number of ranges.
Table~\ref{tbl:writeInval} shows the different types of deletes.
For each, it shows the maximum number of deleted keys and ranges cleared.
$L$ is the number of leaf vertices impacted by a gRW-Tx.
With the last two changes and a supernode, $L$ may be a large value and cause its corresponding gRW-Tx to throw a timeout exception 
%require a long time to complete and throw exceptions 
such that it never completes.
This scenario does not exist in our workload.  Should it happen, 
%it may cause FDB to throw a timeout exception.  
Graph-QP may re-try and count the repeated exceptions.  Once it reaches a threshold, it has detected the supernode scenario.  In this case, it may rely on the application logic to (a) isolate the supernode, (b) delete its edges in batches as a sequence of transactions, followed by (c) deleting the vertex itself as a final transaction.

\subsection{Sub-Query Template Life-Cycle}\label{subsec:lifecycle}
{

%\color{blue}
%Graph Query Processors (Graph-QPs) are stateless virtual servers managed by a service coordinator (SC).
%They are created on demand.
%A newly launched Graph-QP registers itself with SC.
%SC facilitates Graph-QP discovery, removes bad Graph-QPs from a deployment, and dynamically re-sizes a deployment based on the imposed workload. 
%Graph nodes are managed by a service coordinator (SC). When a new graph node is created, it registers itself with SC. SC is used to manage the registered graph nodes, including node discovery, marking down bad nodes, or serving as a centralized place to make decisions based on node resource utilizations. We use SC to manage sub-query template registrations.
A system administrator may add a sub-query template or remove one in an online manner without compromising consistency or restarting the system.
A challenge is how to enable or disable a sub-query template across all Graph-QPs instantaneously in the presence of network delays and different forms of failures.
%This is not trivial since messaging delays and different form of failures prevent enabling or disabling a sub-query template on all Graph-QPs instantaneously. %Figure~\ref{fig:cache_template_lifecycle} shows the life-cycle of a sub-query template. 
To address this challenge, we require four possible states for a sub-query template: \emph{registered}, \emph{installed}, \emph{enabled}, or \emph{removed}. When an admin registers a new sub-query template, its initial state is \emph{registered}. SC initiates a two-phase workflow to enable the sub-query template.
In Phase 1, SC notifies all Graph-QPs to register the new sub-query template.
Once all Graph-QPs in the deployment confirm,
SC changes the state of this sub-query template to \emph{installed}. 
Each Graph-QP upon receiving the register message starts to delete cache entries with gRW-Txs that impact the result of sub-query instances of the template.
These gRW-Txs may delete non-existent cache entries as reads have not yet been enabled and are not populating the cache.
%to be aware of the new sub-query templates (i.e., writes attempt to invalidate/refill cache keys belonging to the sub-query template although there are no cache entries of the sub-query template being produced yet because reads have not been enabled). 

%\begin{figure}[!ht]
%    \centering
    %\includegraphics[width=\linewidth]{figs/SubQuery Template Life-Cycle.pdf}
    %\caption{Sub-Query Template Life Cycle}
    %\label{fig:cache_template_lifecycle}
%\end{figure}

In Phase 2, SC sends a message to all Graph-QPs to activate reads to cache the result of sub-query instances corresponding to the template.
Once all Graph-QPs acknowledge,
%sends a second request to each graph node to switch reads to be aware of the new sub-query template. Similarly, once all nodes acknowledge, 
SC changes the state of the sub-query template to \emph{enabled}. 

The system administrator may disable the caching of a sub-query template.
This may be due to a change in the traffic pattern where reads reference the sub-query template either rarely or never.
In this case,
the administrator does not want the gRW-Txs to incur the overhead of deleting
cache entries that provide no benefit.
%corresponding to the template.
%In case a sub-query template is no longer needed (e.g., the traffic pattern changed and the sub-query template is not being used by queries), 
The administrator uses the SC to initiate the reverse process to remove an enabled sub-query template and to reclaim the space occupied by its cache entries.
This also consists of a two-phase workflow.
In Phase 1, the SC requests each Graph-QP to stop using the sub-query template with reads.
The Graph-QPs continue to require gRW-Txs to delete cache entries corresponding to the instances of this sub-query template.
Once all Graph-QPs confirm that they have stopped reads from using the cache with this sub-query template, the SC 
%and waits until all nodes confirm that the cache-population queues were drained, i.e. all nodes stopped populating new cache entries. 
%Writes, however, still need to be aware of the impacted cache keys. SC 
changes the state of the template to \emph{installed}.
Next, it triggers Phase 2 by requesting each Graph-QP to stop requiring writes to delete cache entries corresponding to the sub-query instances of the disabled sub-query template. Once all Graph-QPs confirm, the SC deletes the cached keys by issuing a clearRange() operation on the sub-query template cached-key prefix, e.g., SQ1 in Figure~\ref{fig:sub_query_template_example}.
This frees up the space occupied by the cached entries.
Next, the SC marks the sub-query template as \emph{removed}.
The SC continues to track deactivated sub-query templates 
%Removed sub-query templates are still being recorded in SC 
for auditing and troubleshooting purposes.

%At each phase, SC must wait for all Graph-QPs to confirm. 
SC waits for all Graph-QPs to confirm a phase.
If its request to a Graph-QP fails or times out, the SC re-sends the request until it receives a successful reply. 

% There are multiple SC replicas where only one primary SC replica is granted with a lease to implement the workflow and modify the state of different query templates.
% The secondary SCs maintain a copy of the current state of the query templates in their metadata database. If the primary SC fails or becomes network unreachable, a secondary replica is assigned the new lease once the lease granted to the primary expires.
% This replica becomes the primary and resumes processing of the workflow. It resets the unfinished phase and resumes the workflow from it.

%}

\subsection{Query Rewriting and Amdhal's Law}\label{sec:rewrite}
Certain gR-Txs retrieve a large amount of data, and/or incur significant Graph-QP processing time.
The former may consist of additional steps that retrieve a large amount of data from FDB.
An example of the latter is merging two large sets without sorting them first.
Both may reference a one-hop sub-query template.
%These queries may reference a one-hop sub-query template and observe a cache hit for it.
A key question is what is the benefit of the cache on these queries?
%in improving the performance of the one-hop sub-query steps of these gR-Txs while leaving their other steps the same?
Amdahl's law~\cite{amdahl67} provides an answer:
%\begin{equation}
$S = \frac{1}{(1-f)+\frac{f}{k}}$.
%\end{equation}
Where $S$ is the effective speedup, $f$ is the fraction of work that is being sped up, and $k$ is the speedup while in the faster mode.
Assume the system spends $f$=90\% of its time processing the one-hop sub-query instances of a gR-Tx.
If the cache speeds up these steps $k$=10x then Amdahl's law predicts the overall gR-Tx speedup to be 5x.
If only $f$=10\% of the gR-Tx time is spent on processing the one-hop sub-query instances then 
Amdahl's law predicts less than 10\% speedup.
Steps that were not improved diminish the cache's 10x improvement considerably, potentially to an insignificant percentage.

%We apply two query re-writing rules to improve the other steps in a gR-Tx.
%The first re-writes the filter step of a gR-Tx to use built-in unique vertex ids instead of user-defined properties with unique values.
We apply re-writing rules to improve those steps of a gR-Tx that do not reference one-hop sub-query instances.  
For example, one of our rules re-writes
%We now describe one of our optimization rules. It re-writes 
the filter step of a gR-Tx to use built-in unique vertex ids instead of user-defined properties with unique values.
Given a vertex id, fetching its property value requires one network round-trip.  The available unique vertex ids eliminates this overhead.
To illustrate, consider a user-defined listing id A with a query to compute all WatchLists that have A and, for each WatchList, retrieve all other listings that are different from A.  
In this case, JanusGraph-generated vertex ids are sufficient to filter listings that are different from the original listing A.
This eliminates one network round-trip time and frees system resources for processing other concurrent requests.
%This eliminates one network round-trip time to fetch the value of their user defined properties with unique values.  It also frees system resources for processing other concurrent requests.

%The second rule rewrites queries that merge two large sets consisting of $m$ and $n$ elements by sorting each set first.  We are interested in large values of $m$ and $n$ because our objective is to reduce the 95$^{th}$ and 99$^{th}$ latency.  Without sorting, the complexity of computing intersection is $O(m \times n)$.  With sorting, the complexity is $O(M \times log~M)$ where $M=max(m,n)$.  It is an improvement with large values of $m$ and $n$.  If min($m$, $n$) < $log~M$ then sorting may be worse. 

\section{Evaluation}\label{sec:eval}
We evaluated our implementation of the one-hop sub-query result cache by mirroring the real-time traffic from our Production system to a Test system.
The Test system implements the one-hop sub-query result cache (Section~\ref{sec:impl}) and query re-writing (Section~\ref{sec:rewrite}).
It provides knobs to enable or disable the cache, denoted C$^+$ and C$^-$, respectively.
Similar knobs enable or disable query re-writing, Q$^+$ and Q$^-$, respectively.
These online knobs support four possible configurations of the Test system:
\begin{enumerate}
  \item \textbf{C$^-$Q$^-$}: Disable cache and disable query re-writing.
  \item \textbf{C$^-$Q$^+$}: Disable cache and enable query re-writing.
  \item \textbf{C$^+$Q$^-$}: Enable cache and disable query re-writing.  
  \item \textbf{C$^+$Q$^+$}: Enable cache and enable query re-writing.
\end{enumerate}

\begin{figure}
    \centering
    \includegraphics[width=\linewidth]{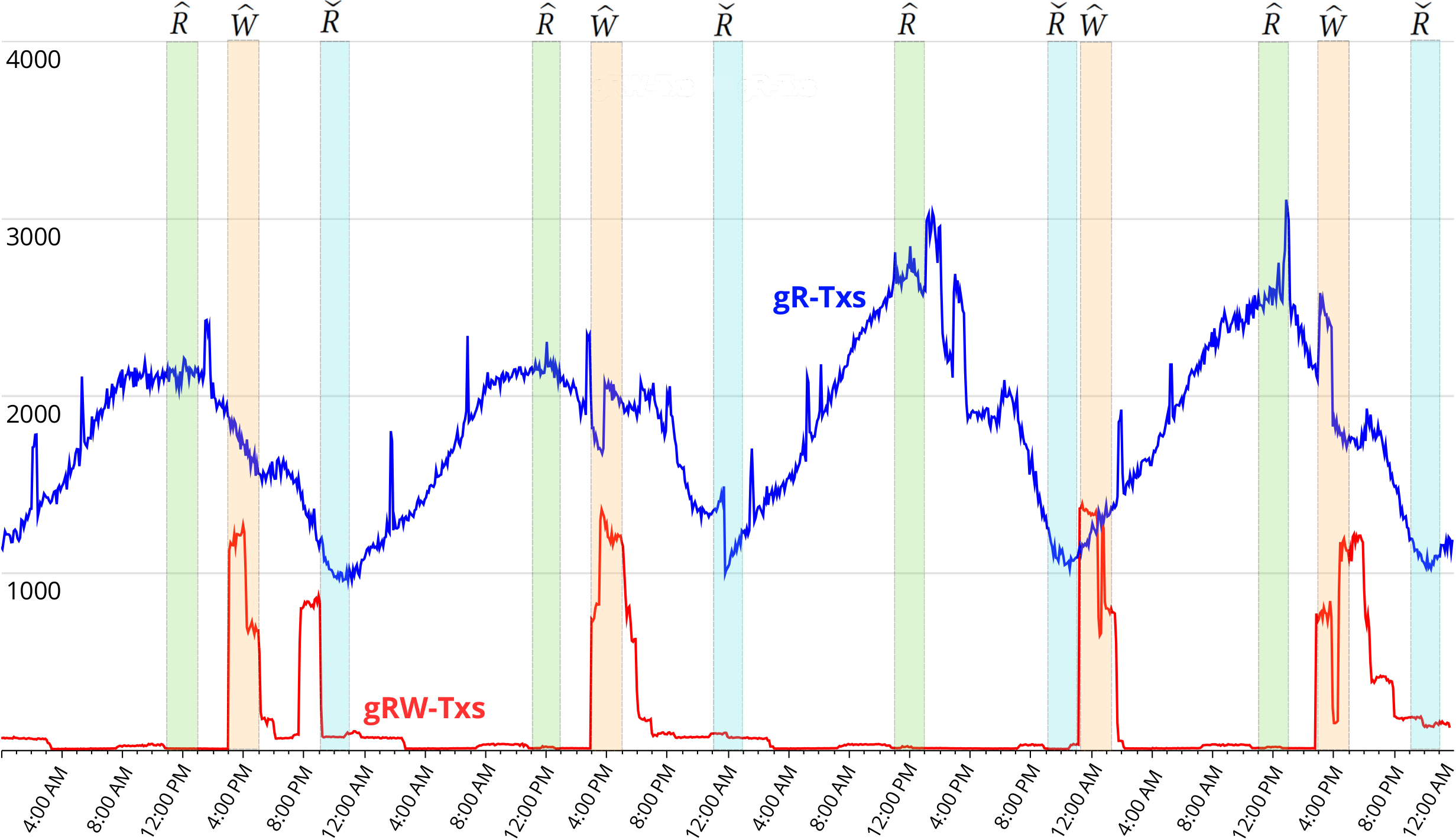}
    \caption{Production traffic for 4 days, primary data center.}
    \label{fig:traffic_pattern}
\end{figure}

Figure~\ref{fig:traffic_pattern} shows our production workload is diurnal and consists of three distinct loads:
\begin{enumerate}
    \item $\widehat R$: A high system load of approximately 2k-3k gR-Txs per second in the primary data center from approximately 11 am to 1 pm.
    \item $\widehat W$: The high system load of gRW-Tx that typically starts at 3 pm %from 3-5 pm 
    daily during an offline batch write process and %starts at 3 pm and 
    may last for approximately 4-12 hours.
    \item $\widecheck R$: A low system load of approximately 1k gR-Txs per second typically from 9-11 pm.    
\end{enumerate}
Query dominated workloads, $\widehat R$ and $\widecheck R$, include approximately 30 gRW-Txs per second.
With $\widehat R$, the mix of gR-Txs and gRW-Txs is 99\% and 1\%.
This mix increases in favor of gRW-Txs with $\widecheck R$, approximately 94\% gR-Txs and 6\% gRW-Txs.
The $\widehat W$ workload consists of approximately 62\% gR-Txs and 38\% gRW-Txs.

This section compares the performance of the four alternative configurations of the Test system with one another for each workload $\widehat R$, $\widehat W$, and $\widecheck R$.
Obtained results highlight the following lessons:
\begin{enumerate}
\item\label{lesson:origQuery} The one-hop sub-query result cache enhances the 95$^{th}$ and 99$^{th}$ percentile latency of gR-Txs several folds with all three system loads.
This is true with both the original and the re-written gR-Txs.
Section~\ref{sec:lesson1}.

%\item The one-hop sub-query result cache in combination with query re-writing enhances the 95$^{th}$ and 99$^{th}$ percentile latency of gR-Txs several folds with all three system loads.
%This was highlighted in Table~\ref{tbl:highlight} of Section~\ref{sec:intro}.
%We discuss this lesson in combination with Lesson 2 in Section~\ref{sec:lesson2}.
%Section~\ref{sec:lesson1} provides additional details.

%\item\label{lesson:origQuery} The cache improves the performance of both the original and re-written gR-Txs.  This is true with all three production workloads. Section~\ref{sec:lesson2}.

\item\label{lesson:writes} Both the 95$^{th}$ and 99$^{th}$ latency of gRW-Txs is improved significantly when the cache is enabled.
Section~\ref{sec:lesson2}.

\item\label{lesson:otherQueries} Queries that do not reference a one-hop sub-query result cache observe a significant enhancement in their 95$^{th}$ and 99$^{th}$ latency when the cache is enabled.
Section~\ref{sec:lesson3}.

\item The cache reduces the percentage of incurred FDB exceptions, enabling a higher percentage of gR-Txs and gRW-Txs to complete successfully.
%with query re-writing (Q$^+$), cache (C$^+$), and both query re-writing and cache (C$^+$Q$^+$).
Section~\ref{sec:lesson4}.
%These techniques reduce the error rate of our transactional storage manager, FDB, significantly. 
\end{enumerate}
%We were surprised with Lessons~\ref{lesson:writes} and~\ref{lesson:otherQueries}.
%This was especially true with Lesson~\ref{lesson:writes} since, with the cache enabled, gRW-Txs must perform extra work by identifying the impacted cache entries and invalidating them.
Below, we provide an overview of our Production and Test system.
%in Section~\ref{sec:TestSystem}.
%Section~\ref{sec:rewrite} presents our query re-writing technique.
Subsequently, Sections~\ref{sec:lesson1}-\ref{sec:lesson4} detail each lesson and its experimental results.

\subsection{Production and Test Systems}\label{sec:TestSystem}
Our production system is deployed across three data centers using FDB's asymmetric configuration model~\cite{fdb7,fdbconfig} for high availability.
%A primary and a secondary data center forms a two-region replication of a FDB cluster~\cite{}.
A primary and a secondary data center maintain a replica of the graph database, see Section 3 of~\cite{fdb_sigmod_2021}.
Each constructs three copies of data. 
The third data center hosts the transaction log records only.

We mirrored the real-time traffic from our Production system to a Test system to quantify the tradeoffs associated with our one-hop sub-query result cache.
Both systems have the same hardware.
Each consists of 288 FDB nodes.
The Test system consists of 180 Graph-QP nodes deployed in the primary data center. 
%Each FDB storage node consists of 3 cores and 18 GB of memory.
%It runs 2 FDB processes.
%Each Graph-QP node consists of 3 cores and 16 GB of memory.

We fork gR-Txs and gRW-Txs from the Production system by configuring its Graph-QP to place them
in a queue while processing them normally.
In the background, these queued requests, including both gR-Txs and gRW-Txs, are forwarded to the Graph-QP nodes of the Test system for processing, see Figure~\ref{fig:mirrored}.

\begin{figure}
\centering
\includegraphics[width=0.9\linewidth]{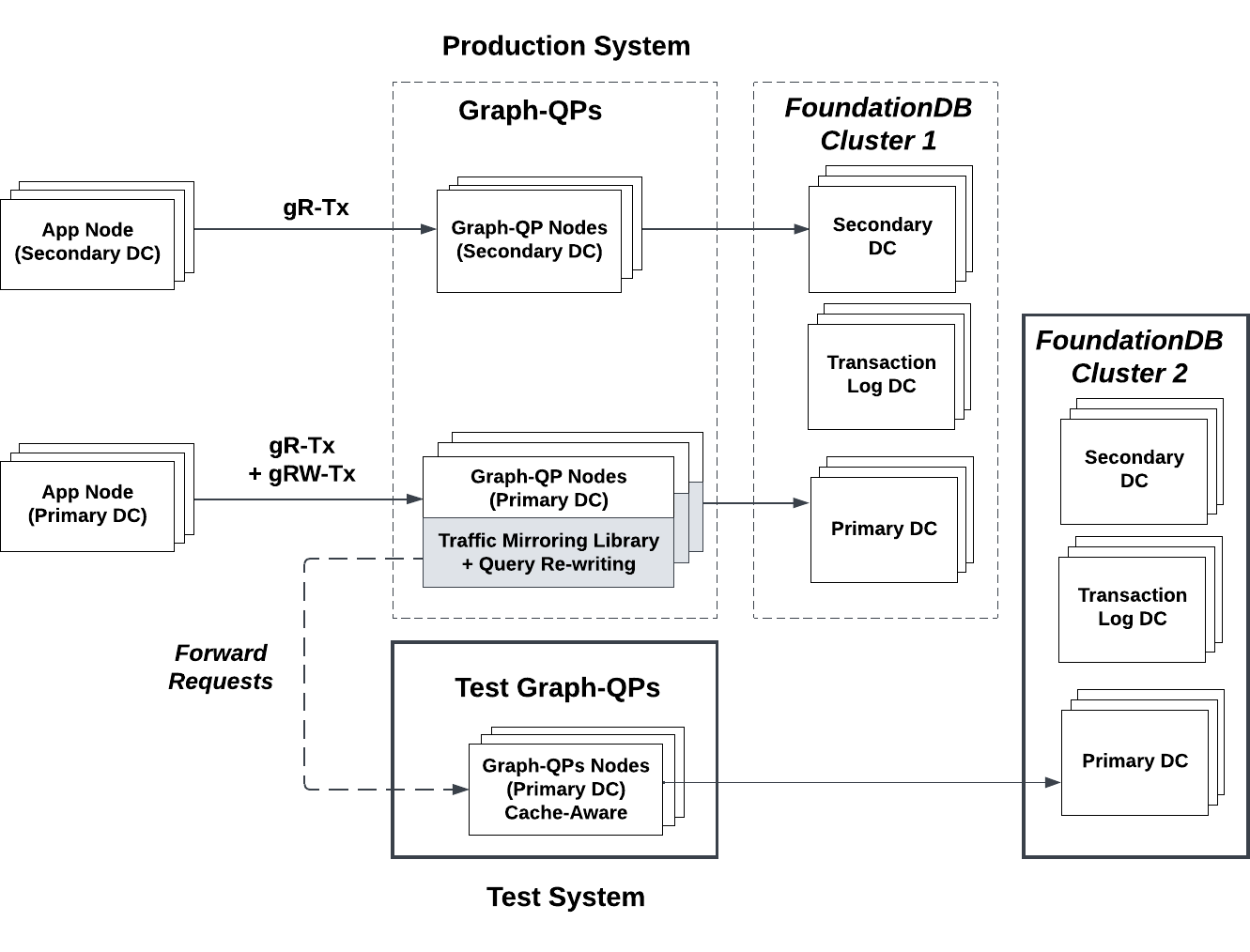}
\caption{Production and Test systems.}
     \label{fig:mirrored}
\end{figure}

An experiment starts by enabling the Test system's logging capability and setting its knobs to a desired configuration (say C$^+$Q$^+$) at a specific time of day that represents one of our three workloads.
Each experiment is 30-minutes in duration.  
We process the resulting log files by discarding their first 5-minutes, processing the remaining 25-minutes to obtain the reported system characteristic including its 95$^{th}$ and 99$^{th}$ latency.

\subsection{Lesson 1: Cache Reduces gR-Tx Latency}\label{sec:lesson1}
\begin{table}
    \centering
    \begin{tabular}{|c||ccc|ccc|}
    \hline
        Work- & \multicolumn{3}{c|}{95$^{th}$ Latency} & \multicolumn{3}{c|}{99$^{th}$ Latency} \\
        \cline{2-7}
         load & C$^-$Q$^+$ & C$^+$Q$^-$ & C$^+$Q$^+$ & C$^-$Q$^+$ & C$^+$Q$^-$ & C$^+$Q$^+$ \\ \hline
        $\widehat R$ & 1.10 & 2.06 & 3.17 & 1.36 & 1.63 & 4.90 \\ \hline
        $\widehat W$ & 1.17 & 2.04 & 2.59 & 1.59 & 1.72 & 4.66 \\ \hline
        $\widecheck R$ & 1.16 & 2.31 & 3.70 & 1.33 & 1.87 & 5.21 \\
    \hline
    \end{tabular}
    \caption{Factor of improvement in the 95$^{th}$ and 99$^{th}$ latency averaged across all gR-Txs.}
    \label{tab:graph_queries}
\end{table}

% \vspace{-10pt}

This lesson was highlighted in Table~\ref{tbl:highlight} of Section~\ref{sec:intro}.
Table~\ref{tab:graph_queries} shows the factor of improvement in the 95$^{th}$ and 99$^{th}$ latency of gR-Txs with all possible settings of the Test system and different workloads.
Our workload consists of approximately 150 unique query templates.
We excluded the query templates that contributed less than 1\% traffic.
With the remaining query templates, we identified six one-hop sub-query templates for caching\footnote{Some of the 1\% ignored queries reference one or more of these one-hop sub-query templates.}.
These cover approximately 36\% of the 150 query templates and approximately 68\% of the gR-Tx traffic to the production system.
We have warmed up the cache for approximately two weeks.
The cache hit rate for the different one-hop sub-query templates ranges from 80\% to 93\%.

Table~\ref{tab:graph_queries} highlights several observations.
First, the one-hop sub-query result cache is more effective than the re-writing rules, compare Columns C$^-$Q$^+$ and C$^+$Q$^-$ of Table~\ref{tab:graph_queries} for each of the 95$^{th}$ %latency and Columns C$^-$Q$^+$ and C$^+$Q$^-$ for the 
and the 99$^{th}$ latency.
Second, the cache improves the performance of 
%both 
the original queries and the re-written queries,
compare Columns $C^{+}Q^{-}$ and $C^{+}Q^{+}$ of Table~\ref{tab:graph_queries} for each of the 95$^{th}$ %latency and Columns C$^+$Q$^-$ and C$^+$Q$^+$ for 
and the 99$^{th}$ latency.

Table~\ref{tab:graph_queries} also highlights the discussion of Amdhal's law in Section~\ref{sec:rewrite}.
The factor of improvement is higher when we speed up both the one-hop sub-query portion of a query (using the cache) and the remaining portion of a gR-Tx (using query re-writing), see C$^+$Q$^+$ column of Table~\ref{tab:graph_queries}.

\begin{table}[!ht]
    \centering
    \begin{tabular}{|c|c||c||c|}

        \hline
        Workload & Experiment & 95$^{th}$ Latency & 99$^{th}$ Latency \\
 %   \cline{3-6}
 %       & Day & C$^+$Q$^-$ & C$^+$Q$^+$ & C$^+$Q$^-$ & C$^+$Q$^+$ \\ 
 \hline 
 \hline

\multirow{3}{*}{%\rotatebox[origin=c]{90}{$\widehat R$}
$\widehat R$} & Day 1 & 10.33  & 1.54  \\
           & Day 2 & 9.09  & 1.75  \\
           & Day 3 & 12.71 & 1.76  \\
%           & Day 4 & 13.00 & 1.73  \\
\hline 

\multirow{3}{*}{%\rotatebox[origin=c]{90}{$\widehat W$}
$\widehat W$} & Day 1 & 10.33 & 1.54  \\
           & Day 2 & 9.43 & 1.46  \\
           & Day 3 & 6.43 & 1.35  \\
%           & Day 4 & 8.83  & 1.37  \\
\hline

\multirow{3}{*}{%\rotatebox[origin=c]{90}{$\widecheck R$}
$\widecheck R$} & Day 1 & 1.86  & 1.90  \\
           & Day 2 & 1.50 & 1.79  \\
           & Day 3 & 1.60 & 1.72  \\
%           & Day 4 & 2.00 & 2.00 & 1.73 & 1.79 \\
\hline    

%\hline
    \end{tabular}
    \caption{Factor of improvement in 95$^{th}$ and 99$^{th}$ latency of a gR-Tx with the cache enabled C$^+$.}
    \label{tbl:707ceecd}
\end{table}

For a gR-Tx, the system load may impact the observed factor of improvement dramatically.
This is shown in Table~\ref{tbl:707ceecd} with one query.
This query does not benefit from query re-writing, i.e., its factor of improvement with query re-writing is approximately 1.
Hence, we show the factor of improvement in its 95$^{th}$ and 99$^{th}$ latency with cache enabled C$^+$ across three different days.
The cache improves this query's 95$^{th}$ latency dramatically with a high system load, both $\widehat R$ and $\widehat W$.
The improvement is significant with a low system load, $\widecheck R$.  
However, not as dramatic as $\widehat R$ and $\widehat W$.
It is interesting to note that the percentage improvement observed with this query's 99$^{th}$ latency and a low system load, $\widecheck R$, is generally higher than that with a high system load.

\begin{figure}[htbp]
    \centering
    % First chart (on the left)
    \begin{subfigure}{0.49\linewidth}  % Use 0.45\linewidth to fit in a single column
        \centering
        \includegraphics[width=\linewidth]{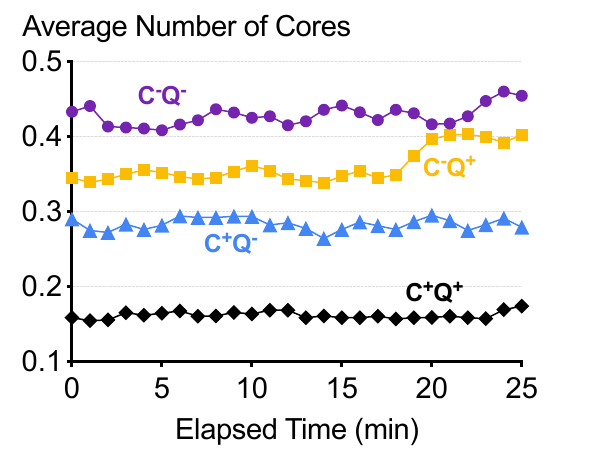}  % Path to the first PDF
        \caption{Graph-QP CPU Utilization}\label{fig:graph-qp-util}
    \end{subfigure}
    % \hspace{0.05\linewidth}  % Space between the two charts
    % Second chart (on the right)
    \begin{subfigure}{0.49\linewidth}
        \centering
        \includegraphics[width=\linewidth]{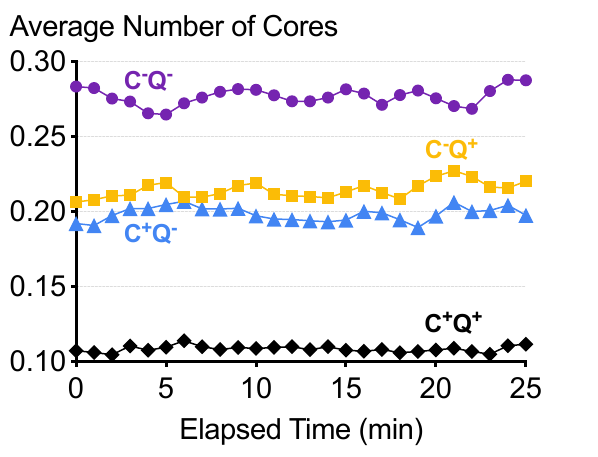}  % Path to the second PDF
        \caption{FDB Client Network Thread}\label{fig:fdb-thread-util}
    \end{subfigure}

    \caption{Graph-QP CPU Utilization with $\widehat R$.}
    \label{fig:graphQPutil}
\end{figure}

The cache reduces the overhead of query processing.
It frees resources to enhance the performance of gRW-Txs and gR-Txs that are not able to use the cache, see Sections~\ref{sec:lesson3} and~\ref{sec:lesson4} for details.
%Figure~\ref{fig:graphQPutil} show the CPU utilization of a Graph-QP node with $\widehat R$.
A Graph-QP node consists of a single threaded FDB client component that processes network I/O to the FDB cluster.
A lower network traffic results in a lower CPU utilization for this thread.  
%Its CPU utilization is shown in Figure~\ref{fig:fdb-thread-util}. 
%It consists of one thread and processes network I/O to the FDB cluster.
Figure~\ref{fig:fdb-thread-util} shows
the CPU consumption of this thread is reduced dramatically with $C^{+}Q^{+}$.
In turn, this reduces the overall CPU utilization of the Graph-QP instance significantly, see Figure~\ref{fig:graph-qp-util}.

\begin{figure*}[htbp]  % htbp defines position as here, top, bottom, page

    % Group 2 (Middle: 3 Charts)
    \begin{subfigure}{\linewidth}
        \centering
        \begin{subfigure}{0.32\linewidth}
            \centering
            \includegraphics[width=\linewidth]{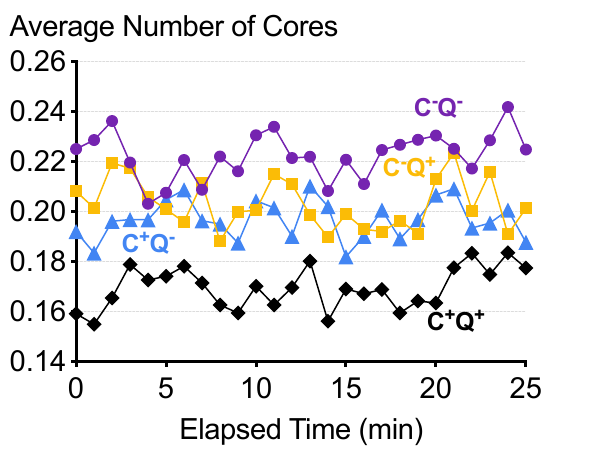}
        \end{subfigure}
        % \hspace{0.03\linewidth}
        \begin{subfigure}{0.32\linewidth}
            \centering
            \includegraphics[width=\linewidth]{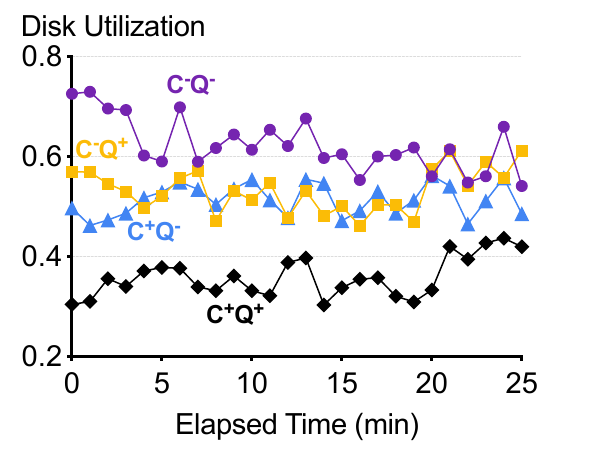}
        \end{subfigure}
        % \hspace{0.03\linewidth}
        \begin{subfigure}{0.32\linewidth}
            \centering
            \includegraphics[width=\linewidth]{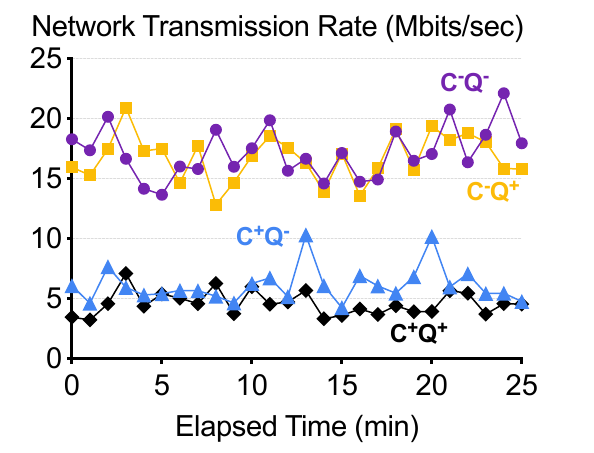}
        \end{subfigure}
        \caption{FDB storage nodes}
        \label{fig:fdb_storage_utilization}
    \end{subfigure}

    \vspace{1em}  % Space between groups

    % Group 3 (Bottom: 3 Charts)
    \begin{subfigure}{\linewidth}
        \centering
        \begin{subfigure}{0.32\linewidth}
            \centering
            \includegraphics[width=\linewidth]{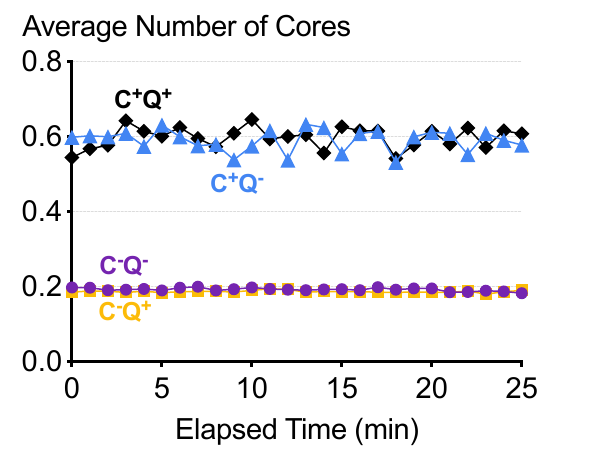}
        \end{subfigure}
        % \hspace{0.03\linewidth}
        \begin{subfigure}{0.32\linewidth}
            \centering
            \includegraphics[width=\linewidth]{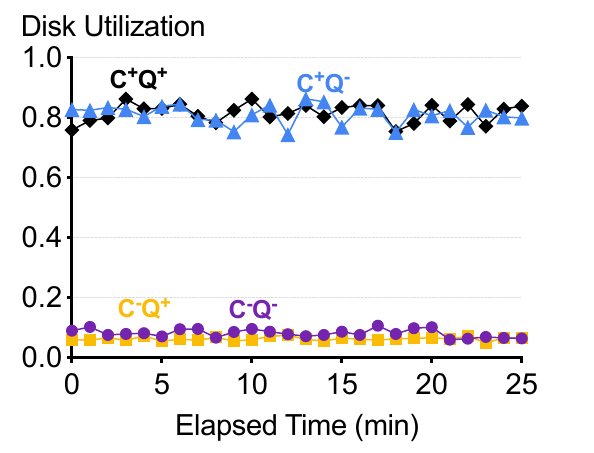}
        \end{subfigure}
        % \hspace{0.03\linewidth}
        \begin{subfigure}{0.32\linewidth}
            \centering
            \includegraphics[width=\linewidth]{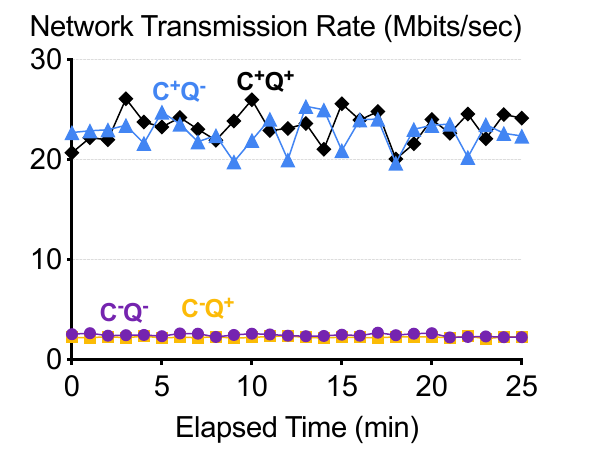}
        \end{subfigure}
        \caption{FDB transaction nodes}
        \label{fig:fdb_xact_node_utilization}
    \end{subfigure}

    \caption{CPU, Disk, and Network utilization of FDB storage and transaction nodes with $\widehat R$.}
    \label{fig:fdb_read_utilization}
\end{figure*}

By reducing the amount of data retrieved from FDB, the cache reduces the CPU, disk, and network utilization of FDB storage nodes.
These stateful nodes maintain the key-value pairs that correspond to our graph database.
Figure~\ref{fig:fdb_storage_utilization} shows the CPU and disk utilization, and network transmission rate of these nodes. 
The cache frees these resources by looking up the result of one-hop sub-queries instead of processing them. 

Figure~\ref{fig:fdb_read_utilization} shows the cache increases the average CPU and disk utilization, and network transmission rate of FDB transaction nodes.
These nodes maintain FDB's transaction log.
With the cache enabled, the sub-queries that observe a cache miss cause the background CP threads to compute cache entries and insert them in FDB in a transactional manner.
This uses FDB's transaction nodes.
Figure~\ref{fig:fdb_write_utilization} shows this overhead of the cache in the form of a higher CPU utilization (3x), disk utilization (more than 4x), and network traffic (almost 5x).
These do not slow down the performance of reads because our design populates the cache asynchronously, utilizing FDB's separation of read and write paths intelligently.
%With FDB, its read path is separate from its write path.  
%Correspondingly, our implementation of a sub-query cache maintains this separation.
%And, our design and implementation of a sub-query cache maintains this separation.

\subsection{Lesson 2: Cache Reduces gRW-Tx Latency}\label{sec:lesson2}
%Tables 7, 13, 14, Figure 12 of the original eval.
A gRW-Tx must identify its impacted cached keys and delete them.
A surprising find is the improvement in the 95$^{th}$ and 99$^{th}$ latency of gRW-Txs instead of a slow-down, see Table~\ref{tab:gRW-Txs}.
There are several reasons for this observation.
First, more than 25\% of the gRW-Txs impact no cache entries because their graph write does not impact the result of a one-hop sub-query template.
An example is a gRW-Tx that modifies the value of a property not referenced by a one-hop sub-query template.

\begin{table}
    \centering
    \begin{tabular}{|c||ccc|ccc|}
    \hline
        Work- 
        & \multicolumn{3}{c|}{95$^{th}$ latency} & \multicolumn{3}{c|}{99$^{th}$ latency} \\
        \cline{2-7}
         load & C$^-$Q$^+$ & C$^+$Q$^-$ & C$^+$Q$^+$ & C$^-$Q$^+$ & C$^+$Q$^-$ & C$^+$Q$^+$ \\ \hline
        $\widehat R$ & 0.99 & 1.10 & 1.76 & 1.00 & 1.47 & 1.99 \\ \hline
        $\widehat W$ & 1.02 & 1.44 & 1.40 & 1.01 & 1.32 & 1.40 \\ \hline
        $\widecheck R$ & 1.01 & 0.98 & 0.98 & 0.98 & 1.25 & 1.24 \\
    \hline
    \end{tabular}
    \caption{Factor of improvement in the %95$^{th}$ and 99$^{th}$ 
    latency of gRW-Txs.}
    \label{tab:gRW-Txs}
\end{table}

Second, a gRW-Tx deletes a few cached keys.
A gRW-Tx is a read-modify-write (RMW) implemented as a multi-step Gremlin expression.
It performs one of the following three types of graph write operations:
(1) Update or insert (Upsert) a sub-graph consisting of several vertices and edges between them, (2) Update last-seen edges, and (3) Delete edges.
Due to the compliance and regulatory requirements, we do not delete vertices.
It evaluates one or more predicates that must hold true in order for the expression to perform its graph write operation.
For example, Type 1 adds a vertex if one does not exist. 
Otherwise, it may update some existing properties of the vertex.
The predicate of Type 1 evaluates whether its specified vertices and edges already exist and their properties are identical with those specified by the graph write.
If this is the case then it becomes a no-op and impacts no cache entries.

While a read of RMW gRW-Tx may reference a one-hop sub-query template, the gRW-Tx does not use the cache.
%we do not service it using the cache.
This ensures the read observes the partial graph writes of its gRW-Tx.
These reads become faster with the cache enabled due to higher availability of system resources, see discussion of Figure~\ref{fig:graphQPutil}.

\begin{table}[!ht]
    \centering
    \begin{tabular}{|l|l|l|l|l|l|}
        \hline
        Write Type & Mean & 50$^{th}$ & 95$^{th}$ & 99$^{th}$ & Max \\ \hline \hline
        Upsert Sub-Graph & 4.72 & 0 & 18 & 28 & 2516 \\ \hline
        Update Last-Seen & 0 & 0 & 0 & 0 & 0 \\ \hline
        Delete Edges & 0.26 & 0 & 4 & 8 & 32 \\ \hline
    \end{tabular}
    \caption{Impacted cached keys per write type.}
    \label{tbl:write_impacted_cache_keys}
\end{table}

\begin{table}[!ht]
    \centering
    \begin{tabular}{|l|c|c|c|}
        \hline
        \multirow{2}{*}{Write Type} & \% of & \multicolumn{2}{|c|}{Factor of Improvement} \\
        \cline{3-4}
         & Writes & 95$^{th}$ Latency & 99$^{th}$ Latency \\ \hline \hline
        Upsert Sub-Graph & 44.85\% & 1.48 & 1.28 \\ \hline
        Update Last-Seen & 43.94\% & 2.13 & 1.48 \\ \hline
        Delete Edges & 11.22\% & 2.14 & 1.39 \\ \hline
    \end{tabular}
    \caption{%Different types of write with 
    $\widehat W$ and the factor of improvement with the cache.}
    \label{tbl:write_heavy_breakdown}
\end{table}

Table~\ref{tbl:write_impacted_cache_keys} shows the number of impacted cache keys by each type of writes with $\widehat W$.
At the 50th percentile, the number of impacted keys is zero with all three types of writes.  \emph{Update Last-Seen} impacts none of the six sub-query templates.  Upsert impacts the highest number of keys when considering the 95$^{th}$ and 99$^{th}$ percentile.
At the 99$^{th}$ percentile, only a few dozens keys are impacted.
At its extreme, a gRW-Tx impacts 2516 cached keys.
It consists of 3 updates.
One changes the property value of a leaf vertex,
impacting all 2516 keys because (1) the vertex property is referenced by two one-hop sub-query templates,
%a part of two different sub-query templates' predicates, namely, SQ3 and SQ4, 
and (2) this vertex has a large number of edges.% connected to it.

\begin{figure*}[htbp]  % htbp defines position as here, top, bottom, page
    % Group 3 (Bottom: 3 Charts)
    \begin{subfigure}{\linewidth}
        \centering
        \begin{subfigure}{0.32\linewidth}
            \centering
            \includegraphics[width=\linewidth]{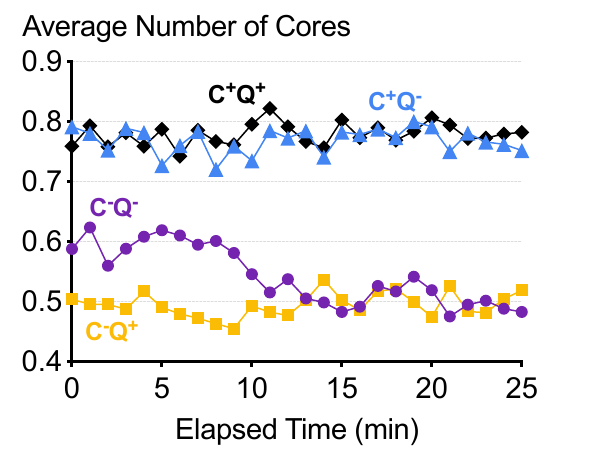}
        \end{subfigure}
        % \hspace{0.03\linewidth}
        \begin{subfigure}{0.32\linewidth}
            \centering
            \includegraphics[width=\linewidth]{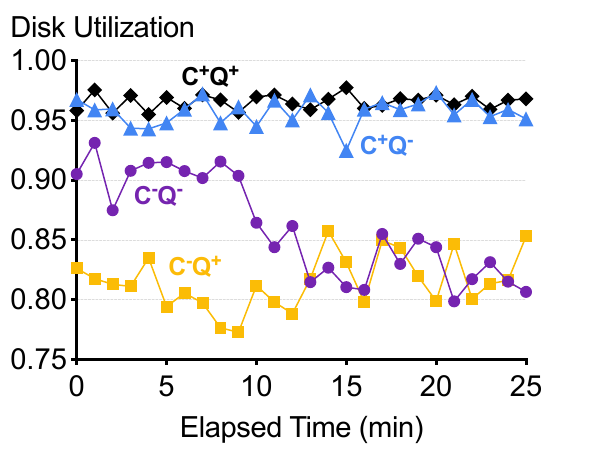}
        \end{subfigure}
        % \hspace{0.03\linewidth}
        \begin{subfigure}{0.32\linewidth}
            \centering
            \includegraphics[width=\linewidth]{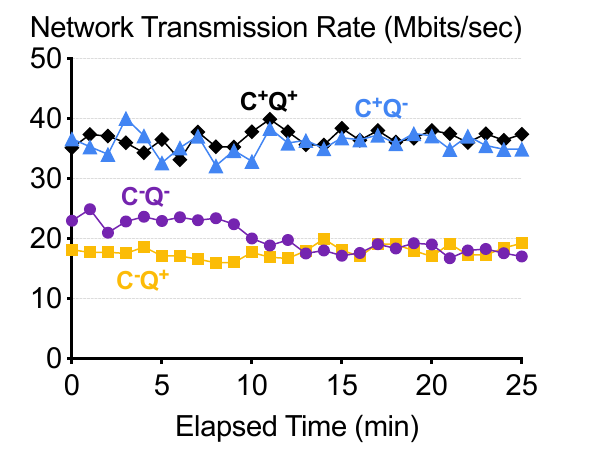}
        \end{subfigure}
        % \caption{FDB transaction nodes}
        \label{fig:fdb_xact_utilization}
    \end{subfigure}

    \caption{CPU, Disk, and Network utilization of FDB transaction nodes with $\widehat W$.}
    \label{fig:fdb_write_utilization}
\end{figure*}

Table~\ref{tbl:write_heavy_breakdown} shows the factor of improvement in the 95$^{th}$ and 99$^{th}$ latency observed with the cache for the three types of graph writes.
The cache improves the performance of their graph reads by reducing the overhead of query processing.
In turn, this reduces the imposed load on the Graph-QPs and FDB instances.  It frees resources to provide the significant improvements shown in 
Table~\ref{tbl:write_heavy_breakdown}.

Third, with the cache disabled (C$^-$), gRW-Txs continue to delete impacted cache entries to maintain it consistent with the graph database.
This enables subsequent experiments using the cache.
However, it increases the latency of gRW-Txs with C$^-$.

With $\widehat W$, Figure~\ref{fig:fdb_write_utilization} shows the cache increases the CPU and disk utilization of FDB's transaction nodes (log servers), resulting in a disk utilization of 94-96\%.
%The disk utilization with the cache is 94-96\%.
%without the cache, 80-90\% with $C^-Q^-$ in Figure~\ref{fig:fdb_write_utilization}.  
Our application inserts new data and the one-hop sub-queries reference this data.
When they observe a cache miss, asynchronous CP threads populate the cache with new entries, increasing the disk utilization of FDB's log servers.
%These writes increase the disk utilization to 94-96\% with $C^+Q^+$ in the same figure.
%We tried to reduce the disk utilization by increasing the number of transaction nodes 2x and 3x.
We increased the number of transaction nodes 2x and 3x.
However, the disk utilization remains unchanged because of FDB's design that requires log records and their meta-data to be written to all log servers, see Section 2.4.3 of~\cite{fdb_sigmod_2021}.
Surprisingly, this higher disk utilization does not degrade the response of writes with $C^+Q^+$.

\begin{table}[!ht]
    \centering
    \begin{tabular}{|c|c||c|c||c|c|}

        \hline
        \multirow{2}{*}{Workload} & Experiment & \multicolumn{2}{c||}{{95$^{th}$ latency}} & \multicolumn{2}{|c|}{{99$^{th}$ latency}} \\
    \cline{3-6}
        & Day & C$^+$Q$^-$ & C$^+$Q$^+$ & C$^+$Q$^-$ & C$^+$Q$^+$ \\ \hline \hline

\multirow{3}{*}{%\rotatebox[origin=c]{90}{$\widehat R$}
$\widehat R$} & Day 1 & 4.54 & 1.43 & 1.68 & 1.77 \\
           & Day 2 & 2.60 & 2.20 & 1.50 & 1.48 \\
           & Day 3 & 2.75 & 2.60 & 1.44 & 1.52 \\
%           & Day 4 &  &  &  &  \\
\hline 

\multirow{3}{*}{%\rotatebox[origin=c]{90}{$\widehat W$}
$\widehat W$} & Day 1 & 7.83 & 7.50 & 1.59 & 1.54 \\
           & Day 2 & 7.00 & 11.75 & 1.50 & 1.59 \\
           & Day 3 & 7.60 & 4.33 & 1.47 & 1.73 \\
%           & Day 4 & & & & \\
\hline

\multirow{3}{*}{%\rotatebox[origin=c]{90}{$\widecheck R$}
$\widecheck R$} & Day 1 & 5.09 & 6.22 & 1.72 & 1.72 \\
           & Day 2 & 4.43 & 4.54 & 1.70 & 1.68 \\
           & Day 3 & 6.11 & 6.11 & 1.74 & 1.82 \\
%           & Day 4 & & & & \\
\hline    

%\hline
    \end{tabular}
    \caption{Factor of improvement in 95$^{th}$ and 99$^{th}$ latency of an aggregate query.}
    \label{tbl:q1}
\end{table}

\subsection{Lesson 3: Cache Expedites Queries Not Referencing One-Hop Sub-Queries}\label{sec:lesson3}

The cache and query re-writing free system resources to process other gR-Txs.
This expedites processing of queries that do not reference one-hop sub-queries.
To illustrate, one gR-Tx constitutes approximately 14\% of the query traffic produced by our production system.
It is an aggregate query that does not have a re-write.
In addition, it does not reference a one-hop sub-query template.
Hence, it does not use the cache.
Still, it benefits greatly when the cache is enabled, see Table~\ref{tbl:q1}. 
Its 95$^{th}$ percentile latency improves several folds, at times more than 11x.  Its 99$^{th}$ percentile improves by at least 1.4x.
%These improvements are observed with the cache by itself and the cache with query re-writing enabled.
The cache reduces the overhead of query processing, reducing the load on Graph-QPs and FDB instance, see discussion of Figure~\ref{fig:fdb_storage_utilization}. 
These freed resources expedite the processing of queries that do not reference one-hop sub-query templates.

\subsection{Lesson 4: Cache Reduces Errors}\label{sec:lesson4}

Our workload incurs the following FDB exceptions:
Transaction timeouts, conflicts, and read requests for future versions.
The first two constitute more than 96\% of observed errors with all configurations of the Test system. 
FDB throws a timeout exception when a gR-Tx or gRW-Tx exceeds the 5-second transaction limit enforced by FDB.
It throws a conflict exception when its optimistic concurrency control technique detects a conflict at commit time of a gRW-Tx, aborting the gRW-Tx.  

Table~\ref{tbl:error} shows the total percentage of queries and transactions that observe an exception with different settings of the Test system.
Query re-writing in combination with the cache reduces these exceptions several-fold, see the last 3 columns of Table~\ref{tbl:error}.
%With the query dominated workloads, 
With $\widehat R$ and $\widecheck R$, almost all errors are due to timeouts.
Query re-writing and the cache expedite processing of queries and make resources available to reduce the likelihood of timeouts.

\begin{table}[!ht]
    \centering
    \begin{tabular}{|c|l|l|l|l|l|l|l|}
        \hline
        Work- & \multicolumn{4}{|c|}{\% Error} & \multicolumn{3}{|c|}{Improvement} \\ \cline{2-8}
        load & {\scriptsize C$^-$Q$^-$} & {\scriptsize C$^+$Q$^-$} & {\scriptsize C$^-$Q$^+$} & {\scriptsize C$^+$Q$^+$} & {\scriptsize C$^+$Q$^-$} & {\scriptsize C$^-$Q$^+$} & {\scriptsize C$^+$Q$^+$} \\ \hline\hline

        $\widehat R$ & 0.592\%	& 0.467\% & 0.214\% & 0.088\% & 1.27 & 2.77 & 6.75 \\ \hline
        $\widehat W$ & 0.499\% & 0.363\% & 0.302\% & 0.163\% & 1.38 & 1.65 & 3.06 \\ \hline
        $\widecheck R$ & 0.478\% & 0.381\% & 0.160\% & 0.081\% & 1.25 & 2.98 & 5.93 \\ \hline

    \end{tabular}
    \caption{Cache and query re-writing reduce the percentage error observed from FDB.}
    \label{tbl:error}
\end{table}

With $\widehat W$, when the cache and query re-writing are disabled, approximately 60\% of exceptions are timeouts and 40\% are conflicts.
Once we enable the cache, exceptions attributed to timeouts are reduced to 30\% while the remaining 70\% are due to conflicts.
%It is interesting to note that, with $\widehat W$, query re-writing does not reduce the percentage of conflicts.
%However, the cache reduces these exceptions by approximately 43-53\%. 

%The two major causes of errors are \emph{time-outs} and \emph{write transaction conflicts}. The former is because some queries retrieve and process a large amount of data that exceed the 5-second transaction limit enforced by FDB. The latter is because concurrent write transactions may have read-write conflicts that cause only one to commit while the others abort due to FDB's use of the optimistic concurrency control protocol. Table~\ref{tbl:error_rate} shows the rate of error for the three workload periods. By reducing latency, caching and query rewrites minimize the rate of errors. By enabling only one, the error rate is reduced 1.16-2.8x. With both enabled, the error rate is reduced significantly, 4.98-6.32x. See the last column of Table~\ref{tbl:error_rate}.

%These conflicts are reduced with either query-rewriting or caching as results of reducing latency. Both caching and query-rewriting are effective ways to reduce latency. We observed improvements in error rate when enabling them one by one (1.16-2.8x). Combining both results in significant reductions in error rate in all the three chosen periods (4.98-6.32x).

%\input{eval}

%\section{Discussion}\label{sec:discuss}
%\input{discuss}

\section{Related Work}\label{sec:related_work}
To the best of our knowledge, the concept of caching the result of one-hop sub-queries that constitute a graph query is novel and not described elsewhere. 
Many studies investigate graph data analytics by focusing on workloads consisting of PageRank~\cite{pr2018}, Breadth First Search (BFS), and connected components~\cite{graph2021,xtree2022,shortcutting2018,gemini2016,topox}.
Our work is different because it investigates processing interactive graph queries in an online manner.
The read queries are issued by users who are waiting for their results.
Hence, it is important to expedite their 95$^{th}$ and 99$^{th}$ percentile response times. 

In-memory graph databases~\cite{coro2024,ligra2013} focus on the use of CPU caches~\cite{gpop2019,cacheEff2021}
%, e.g., L1, 
to enhance performance.
Algorithms designed for this type of a cache are fundamentally different than the caching solutions discussed in our study.
Ours are focused on large graph databases that reside on a secondary storage device.

In our application, the concept of transactions with ACID semantics are a requirement.  Unlike social networking~\cite{bytegraph2022,ramptao2021}, our applications do not tolerate a cache that diverges from the state of the database and produces obsolete results.  Stale and inconsistent data are unacceptable for our use cases.

%{\color{blue} 
Facebook memcache~\cite{scalingmemcache2013} uses memcached servers~\cite{writeback2019,leases2014,camp2014} as a look-aside cache, where MySQL servers act as the authoritative source of data. TAO~\cite{tao2013} employs a write-through cache~\cite{ebay2023} where there are separate cache tiers, a leader and multiple followers, and a storage layer. In both cases, cache entries are stored separately from the database. 
%A page rendering request from a user may trigger fetching hundreds of cache entries belonging to multiple snapshots of the database.
The design goal of both systems is to prioritize large throughput (billion reads and million writes per second) over strong consistency. Like memcache and TAO, our solution supports multi-datacenters and multiple copies for cache entries for low latency and high availability.
Unlike Facebook's systems which only offer eventual consistency, our solution provides strong consistency because we store cache entries persistently with graph data under the same transaction manager. While an improvement on TAO~\cite{ramptao2021} supports atomic-write visible on reads, our system supports snapshot isolation. 
%TAO caches individual nodes and edges as cache entries while in our solution, each cache value is the result of a one-hop sub-query instance.

Our concept of one-hop sub-query result cache may be used by distributed graph processing systems such as Gemini~\cite{gemini2016}, GraphScope~\cite{graph2021}, Grasper~\cite{grasper2019,graphit2020} and others.
Below, we describe Grasper and how our work is different and complementary.
%There are several in-memory systems for property graph databases~\cite{grasper2019,grasper2020,graph2021}.
Grasper is an OLAP system for property graph databases~\cite{grasper2019,grasper2020}.
Each of its nodes consist of a data store and a query engine that communicate using RDMA.
It uses a data flow execution paradigm~\cite{gamma1990} to process queries, performing filtering operations as early as possible.  
It identifies sub-queries that can be performed in parallel and uses intra-query parallelism to expedite their processing. % query execution time. 
Our work is similar in that we are also focused on property graph databases and the Gremlin query language~\cite{marko2015}. 
At the same time, our work is different in several ways. 
First, our workload is transaction processing and is user facing.  These transactions must be processed quickly. 
Second, we separate our data store (FDB) from the graph query engine (Graph-QP), see Figure~\ref{fig:arch}.
This enables each to scale up and down independently~\cite{novalsm2021}.
Third, we do not use either RDMA or intra-query parallelism. 
With the latter, we explored the feasibility of extending JanusGraph for intra-query parallelism and decided the resulting implementation would be too brittle,
see Section~\ref{sec:conclude}.% for details.
%Finally, Grasper does not cache the result of sub-queries while the focus of our entire study is on caching the result of one-hop sub-queries and maintaining them consistent with the state of the persistent graph database.

Some systems use in-memory caches at a node to stage the graph vertices and edges for query processing~\cite{ramptao2021,grasper2019,grasper2020}.
%Grasper uses caches at each of its nodes to stage the graph vertices and edges for query processing.
Our cache is different because its entries are the result of one-hop sub-queries, identifying a list of vertices.
%Grasper caches are in-memory while our 
Our cache resides in a persistent data store, FDB.
Arguably our reported performance numbers may be enhanced if the cache is in-memory, see Section~\ref{sec:conclude}.
%Such an in-memory implementation of our one-hop sub-query result cache is orthogonal to Grasper's design decisions.
%Its extensions with RDMA may be used by Grasper to further enhance its performance for repeat OLAP queries. 

In~\cite{ebay2023}, we introduced a non-transparent\footnote{A non-transparent cache requires an application developer to maintain the cached query results consistent with the database.} cache at the granularity of the result of a graph query.
We used this cache with ten distinct queries and improved their latency 4x-50x.
This helped reduce the overall 95$^{th}$ and 99$^{th}$ percentile latency of reads by 1.3x and 1.5x, respectively.
A challenge of this form of caching is how to identify cache entries impacted by writes.
The complexity of such detection is a function of the complexity of the cached graph queries.  
With some queries, it is difficult if not impossible to compute the impacted keys.
While the selected queries were manageable, they slowed writes down 5x-10x~\cite{ebay2023}.

A one-hop sub-query is a simple part of a complex graph query.
Its cached results may be used by multiple queries that are very different.
%With our workload, one one-hop sub-query template (SQ2)
%For example SQ2 
%is used by 15 different queries with varying complexity.
%, see Column 4 of Table~\ref{tbl:query_template_latency}.
With our production workload, six sub-queries enabled almost all queries in our workload to use the cache.
This improves the overall 95$^{th}$ and 99$^{th}$ percentile latency of reads by at least 2x and 1.63x, respectively.
Another novel contribution is the use of query-rewriting.  
Once combined with the cache, it improves the overall 95$^{th}$ and 99$^{th}$ by at least 2.59x and 4.66x when compared with a system that does not use these techniques.
%The writes become faster because of the simplicity of the sub-queries and freed resources.

\section{Conclusion and Future Work}\label{sec:conclude}
This study presents the design and implementation of a one-hop sub-query result cache.
This novel caching solution expedites the processing of Gremlin graph queries while providing strong transactional guarantees.
It is implemented in the graph processing engine JanusGraph and transparent to its users who issue Gremlin queries.
%{\color{blue}
%We presented sub-query result cache, our novel caching solution for graph databases. 
Our evaluation using real production workloads shows that it enhances the 95$^{th}$ and 99$^{th}$ response time of gR-Txs significantly (at least 2.59x and 4.66x with query rewriting, see Table~\ref{tab:graph_queries}). An astounding result is that the performance of both gRW-Txs and gR-Txs that do not consist of one-hop sub-queries is also enhanced, 1.3x-2.1x and 1.47x-11.75x, respectively.
There are two reasons for this.
First, we cache only one-hop sub-queries.
This simplifies the computation of cached keys by gRW-Txs.
%This bounds the numbers of cached keys impacted by a write.
Second, caching %sub-query results 
reduces the overhead of query processing.
This frees resources for use by gRW-Txs and gR-Txs that do not use the cache.

\balance

We are pursuing several short term research directions. 
First, we are extending our transactional storage manager (FDB) to be cache aware, enabling the cache entries to reside in memory.
%We speculate this will further enhance system response time. 
Second, we are investigating the feasibility of a data flow architecture~\cite{gamma1990,grasper2019,a12020,bytegraph2022,dryad2007,banyan2022} that is cache aware and provides strong consistency guarantees.
The key idea is to control intra-query and inter-query parallelism judiciously~\cite{hrps1990,magic1992,copeland1988} with the objective to further enhance the 95$^{th}$ and 99$^{th}$ percentile response times while maximizing the overall system throughput.

\begin{acks}
This research was a collaborative effort between eBay and the University of Southern California, made possible through support from eBay's eRUPT program. We extend our gratitude to USC students Eric Cheng and Zihao Dong for their valuable assistance in developing unit tests. We also thank eBay's Flora Zhang for her essential role in setting up the test system for our evaluation.
\end{acks}

%\clearpage

\bibliographystyle{ACM-Reference-Format}
\bibliography{sample}

\appendix
\section{A Write-Around Implementation}\label{sec:pseudocode}
This appendix provides the pseudo-code for nine algorithms used in our implementation of the write-around policy. These algorithms are incorporated into Graph-QP, as shown in Figure~\ref{fig:arch}, with FoundationDB (FDB) as the storage manager.
These algorithms are categorized into two groups, utility algorithms and those that implement the changes described in Section~\ref{sec:writes}.
The latter depends on the former.
The utility algorithms are Evaluate, DeleteKeysForRoot, DeleteKeysForLeaf, HandleEdgeChange, and ExtractWildcardValues.
The algorithms that use these to maintain the cache consistent with the graph database include Delete Vertex, Change Vertex Property, Add/Delete Edge, and Change Edge Property.
These are triggered by the Gremlin provided mutation listener interface described in Section~\ref{sec:impl}.
Hence, these algorithms only detect and delete the cache entries.
The mutation of the graph database is performed by the Graph-QP.

The algorithms are self explanatory when considered in the context of the material presented in Sections~\ref{sec:writes} and~\ref{sec:impl}.  Below, we highlight some of their interesting design decisions and nuances.

Change Vertex Property, Algorithm~\ref{alg:vertex_property_changed}, handles both deletion and addition of a property as a special case.
The same is true with Change Edge Property, Algorithm~\ref{alg:edge_property_changed}.
Both assume if the input value of $newval$ is \texttt{null} then the property is being deleted.
If $P$ is missing from $V_i$/$E$ and $newval$ has a valid value then the property is being added.  

The utility algorithms DeleteKeysForLeaf and HandleEdgeChange delete keys from the cache one at a time.  They assume FDB's fat client that buffers these deletes and submits them to its server for processing at transaction commit time.  If a thin client is assumed that does not buffer writes then these utility algorithms should be adjusted to identify all keys to be deleted and submit one delete request for all keys to the server component of the storage manager.  

DeleteKeysForRoot demonstrates another feature of FDB, its ability to clear a range of values starting with a prefix.
It simply computes the first portion of a key consisting of the identifier of the query template along with the vertex id as the prefix.  
FDB deletes all keys starting with this prefix.

% Delete a vertex
\begin{algorithm}[H]
\caption{\textbf{Delete Vertex ($V_i$)}}
\begin{algorithmic}[1]
\Statex \textbf{Description:} Triggered when vertex $V_i$ is deleted.
\For{each $SQ_t$ with $P^r$, $P^e$ and $P^l$}
%  \State match\_root := \FuncCall{Evaluate}($P^r$, $V_i$)
  \If{\FuncCall{Evaluate}($P^r$, $V_i$) == \True}
%  \If{match\_root}
    \State \FuncCall{DeleteKeysForRoot}($SQ_{t}$, $V_{i}$)
  \EndIf
  \item[] % To insert a blank line without numbering
%  \State match\_leaf := Evaluate($P^l$, $V_i$)
%  \If{match\_leaf}
%   \If{\FuncCall{Evaluate}($P^l$, $V_i$) == \True}
   \State \FuncCall{DeleteKeysForLeaf}($V_i$, $SQ_t$)
%  \EndIf
\EndFor
\end{algorithmic}
\label{alg:del_vertex}
\end{algorithm}

% Add/Update/Delete a property of a vertex
\begin{algorithm}[H]
\caption{\textbf{Change Vertex Property ($V_i$, $P$, $newVal$)}}
\begin{algorithmic}[1]
\Statex \textbf{Description:} Triggered when the property $P$ of $V_i$ is changed. If $newVal$ is \emph{null}, $P$ is deleted from $V_i$. If $newVal$ is not \emph{null}, $V_i$ has $P$ set to $newVal$.
\State $V_{old}$ $\gets$ $V_i$
\State $V_{new}$ $\gets$ $V_i$ with P = newVal
\For{each $SQ_t$ with $P^r$, $P^e$, and $P^l$}
  \If{P appears in $P^r$}
    \If{\FuncCall{Evaluate}($P^r$, $V_{old}$) \textbf{or} \FuncCall{Evaluate($P^r$, $V_{new}$)}}
      \State \FuncCall{DeleteKeysForRoot}($SQ_t$, $V_i$)
    \EndIf
  \EndIf
  \Statex
  \If{P appears in $P^l$}
    \State \FuncCall{DeleteKeysForLeaf}($V_{old}$, $SQ_t$)
    \State \FuncCall{DeleteKeysForLeaf}($V_{new}$, $SQ_t$)
  \EndIf
\EndFor
\end{algorithmic}
\label{alg:vertex_property_changed}
\end{algorithm}

% Add/Delete an edge from $V_i$ to $V_j$
\begin{algorithm}[H]
\caption{\textbf{Add/Delete Edge ($E$, $V_i$, $V_j$)}}
\begin{algorithmic}[1]
\Statex \textbf{Description:} Triggered when an edge $E$ from $V_i$ to $V_j$ is added/deleted.
\For{each $SQ_t$ with $P^r$, $P^e$, and $P^l$}
  \State \FuncCall{HandleEdgeChange}($SQ_t$, $E$, $V_i$, $V_j$)
\EndFor
\end{algorithmic}
\label{alg:edgeadddelete}
\end{algorithm}

\begin{algorithm}[H]
\caption{\textbf{Change Edge Property ($E$, $V_i$, $V_j$, $P$, $newVal$)}}
\begin{algorithmic}[1]
\Statex \textbf{Description:} Triggered when the property P of edge E from $V_i$ to $V_j$ changes its value. If $newVal$ is \emph{null}, P is deleted from E. If $newVal$ is not \emph{null}, E has P set to $newVal$.
\State $E_{old}$ $\gets$ $E$
\State $E_{new}$ $\gets$ $E$ with $P$ = $newVal$
\For{each $SQ_t$ with $P^r$, $P^e$ and $P^l$}
  \If{$P$ appears in $P^e$}
    \State \FuncCall{HandleEdgeChange}($SQ_t$, $E_{old}$, $V_i$, $V_j$)
    \State \FuncCall{HandleEdgeChange}($SQ_t$, $E_{new}$, $V_i$, $V_j$)
  \EndIf
\EndFor
\end{algorithmic}
\label{alg:edge_property_changed}
\end{algorithm}

%Below are the pseudo-code for the supplemental functional calls.

% Evaluate the graph element (vertex/edge) with the given predicate
\begin{algorithm}[H]
\caption{\textbf{Evaluate} (Pred, X)}
\begin{algorithmic}[1]
\Statex \textbf{Description:} Evaluate the graph element X, either a vertex or an edge, with the given predicate Pred.
\State labelPred $\gets$ Extract label predicate from Pred
\State propPreds $\gets$ Extract property predicates (without wildcards) from Pred

\If{label\_predicate applied to X.label == \False}
  \State \Return \False
\EndIf

\For{each $propP$ \textbf{in} propPreds}
  \If{X.P is missing \textbf{or} $propP$ applied to X.P == \False}
    \State \Return \False
  \EndIf
\EndFor

\State \Return \True
\end{algorithmic}
\label{alg:evaluate}
\end{algorithm}

% Delete all keys given the root vertex id
\begin{algorithm}[H]
\caption{\textbf{DeleteKeysForRoot ($SQ_t$, $V$)}}
\begin{algorithmic}[1]
\Statex \textbf{Description:} Delete all cache keys where root vertex is $V$.
\State $prefix$ $\gets$ Concat $SQ_t$.name with $V$.id
\State Clear range starting with $prefix$ from cache
\end{algorithmic}
\label{alg:delete_keys_for_root}
\end{algorithm}

\begin{algorithm}[H]
\caption{\textbf{DeleteKeysForLeaf (L, $SQ_t$)}}
\begin{algorithmic}[1]
\State $P^r$, $P^e$, $P^l$ $\gets$ root, edge, and leaf predicates of $SQ_t$
\If{L does not have all wildcard properties in $P^l$}
 \State \Return
\EndIf
\If{\FuncCall{Evaluate}($P^l$, L) == \False}:
  \State \Return
\EndIf
\State $W^l$ $\gets$ \FuncCall{ExtractWildcardValues}($P^l$, L)
\Statex
\Statex \COMMENT{Identify the direction to query for potential edges and roots}
\State tDir $\gets$ Edge direction specified in $SQ_t$
\State $reverseDir$ $\gets$ Init direction of the reverse query
\If{tDir == $outgoing$}
  \State $reverseDir$ $\gets$ $incoming$
\ElsIf{tDir == $incoming$}
  \State $reverseDir$ $\gets$ $outging$
\ElsIf{tDir == $both$}
  \State $reverseDir$ $\gets$ $both$
\EndIf
\item[]
\State Es $\gets$ Edges with direction $reverseDir$ from V that satisfies $P^e$
\For{each edge E in Es}
  \If{E does not have all widcard properties in $P^e$}
    \State \Continue
  \EndIf
  \State $W^e$ $\gets$ \FuncCall{ExtractWildcardValues}($P^e$, E)
  \Statex
  \State R $\gets$ Vertex different than L (at the other side) from E
  \If{\FuncCall{Evaluate}($P^r$, R) == True}
    \State $k$ $\gets$ Concat $SQ_t$.name, R.id, $W^e$, $W^l$
    \State Delete $k$ from the cache
  \EndIf
\EndFor
\end{algorithmic}
\label{alg:compute_keys_from_leaf}
\end{algorithm}

% Add/Delete an edge from $V_i$ to $V_j$
\begin{algorithm}[H]
\caption{\textbf{HandleEdgeChange} ($SQ_t$, $E$, $V_i$, $V_j$)}
\begin{algorithmic}[1]
\State $P^r$, $P^e$, $P^l$ $\gets$ root, edge, and leaf predicates of $SQ_t$
\If{E does not have all wildcard properties in $P^e$}
  \State \Return
\EndIf
\If{\FuncCall{Evaluate}($P^e$, $E$) == \False}:
  \State \Return
\EndIf
\State $W^e$ $\gets$ \FuncCall{ExtractWildcardValues}($P^e$, E)
\Statex
\Statex \COMMENT{Compute possible \{root, leaf\} pairs (at most 2)}
\State rootLeafCandidates $\gets$ Init array []
\State direction $\gets$ Edge direction specified in $SQ_t$
\If{direction == $outging$ \textbf{or} $both$}
  \State Append \{$V_i$, $V_j$\} to rootLeafCandidates
\EndIf
\If{direction == $incoming$ \textbf{or} $both$}
  \State Append \{$V_j$, $V_i$\} to rootLeafCandidates
\EndIf
\Statex
\For{each R, L \textbf{in} rootLeafCandidates}
  \If{L does not have all wildcard properties in $P^l$}
    \State \Continue
  \EndIf
  \Statex
  \State matchRoot $\gets$ \FuncCall{Evaluate}($P^r$, R)
  \State matchLeaf $\gets$ \FuncCall{Evaluate}($P^l$, L)
  \If{matchRoot == \True \textbf{ and} matchLeaf == \True}
    \State $W^l$ $\gets$ \FuncCall{ExtractWildcardValues}($P^l$, L)
    \State $k$ $\gets$ Concat $SQ_t$.name, R.id, $W^e$, $W^l$
    \State Delete $k$ from the cache
  \EndIf
\EndFor
\end{algorithmic}
\label{alg:HandleEdgeChange}
\end{algorithm}

% Extract wildcard values from element X and predicate P
\begin{algorithm}[H]
\caption{\textbf{ExtractWildcardValues (Pred, X)}}
\begin{algorithmic}[1]
\Statex \textbf{Description:} Extract values of wildcard properties of the predicate Pred from graph element X, either a vertex or an edge.
\State wildcardVals $\gets$ Init map \{\}
\For{each wildcard property P in Pred}
  \State Add (P.name, X.P) to wildcardVals
\EndFor
\State \Return wildcardVals
\end{algorithmic}
\label{alg:extract_wildcard_values}
\end{algorithm}

% % Merge wildcard values from edge and leaf
% \begin{algorithm}[H]
% \caption{\textbf{MergeWildcardValues}($W^e$, $W^l$)}
% \begin{algorithmic}[1]
% \Statex \textbf{Description:} Merge the wildcard values of edges ($W^e$) and leaves ($W^l$) into a single map
% \State W $\gets$ Init map \{\}
% \For{each \{P, V\} in $W_e$}
%   \State Put (P, V) to W
% \EndFor
% \For{each \{P, V\} in $W_l$}
%   \State Put (P, V) to W
% \EndFor
% \State \Return W
% \end{algorithmic}
% \label{alg:extract_wildcard_values}
% \end{algorithm}

% % Construct cache keys
% % Expect root_vertex_ids and wildcard_vals equal in size
% \begin{algorithm}
% \caption{\textbf{ConstructCacheKeys}($SQ_t$, $rootId$, $W^e$, $W^l$)}
% \begin{algorithmic}[1]
% \State key $\gets$ Concat $SQ_t$.name, $rootId$, and all \{P, V\} in $W^e$ and $W^l$
% \State \Return key
% \end{algorithmic}
% \label{alg:construct_cache_keys}
% \end{algorithm}

\end{document}